\documentclass{article}
\usepackage{graphicx} % Required for inserting images
\usepackage{siunitx}
\usepackage{listings}
\usepackage{hyperref}
\usepackage{nomencl} % If needed
\makenomenclature
\usepackage{amsmath}
\usepackage{algorithm}
\usepackage{algpseudocode}
\usepackage{bm}
\usepackage{booktabs}
\usepackage[letterpaper, total={6in, 9in}]{geometry}
\usepackage{breqn}
\usepackage{orcidlink}

\usepackage{multirow} 
\usepackage{longtable}
\usepackage{minted}

\usepackage{xcolor}

\definecolor{codegreen}{rgb}{0,0.6,0}
\definecolor{codegray}{rgb}{0.5,0.5,0.5}
\definecolor{codepurple}{rgb}{0.58,0,0.82}
\definecolor{backcolour}{rgb}{0.95,0.95,0.92}
\lstdefinestyle{mystyle}{
    backgroundcolor=\color{backcolour},   
    commentstyle=\color{codegreen},
    keywordstyle=\color{magenta},
    numberstyle=\tiny\color{codegray},
    stringstyle=\color{codepurple},
    basicstyle=\ttfamily\small,
    breakatwhitespace=false,         
    breaklines=true,                 
    captionpos=b,                    
    keepspaces=true,                 
    numbers=left,                    
    numbersep=5pt,                  
    showspaces=false,                
    showstringspaces=false,
    showtabs=false,                  
    tabsize=2,
    upquote=false
}

\usepackage{caption}
\usepackage{subcaption}

\newcommand{\maj}{\Sigma_{\scriptstyle\text{maj}}}

\title{Hybrid Delta Tracking Schemes Using a Track-Length Estimator\footnote{Data Available from Zenodo\cite{delta_restuls}} \footnote{This is an Accepted Manuscript of an article published by Taylor \& Francis in The Journal of Computational and Theoretical Transport on February 22nd 2026, available at: \url{https://doi.org/10.1080/23324309.2026.2618791}}.
}
\author{
  \orcidlink{0000-0003-1379-5431}Joanna Piper Morgan$^{1}$\footnote{Contact: morgan83@llnl.gov, joannapipermorgan@gmail.com} \footnote{Currently: Nuclear Criticality Safety Division, Lawrence Livermore National Lab, Livermore, CA, 94550, USA}
  \and
  \orcidlink{0000-0003-3426-7160}Ilham Variansyah$^{2}$
  \and
  \orcidlink{0000-0003-3358-5618}Kayla B.~Clements$^{2}$\footnote{Currently: TerraPower LLC, Bellevue, WA 98008, USA}
  \and
  Todd S.~Palmer$^{2}$
  \and
  \orcidlink{0000-0003-4425-7097}Kyle E.~Niemeyer$^{1}$
}

\date{%
    \small{
    $^1$School of Mechanical, Industrial, and Manufacturing Engineering, Oregon State University, Corvallis, OR, 97331, USA\\%
    $^2$School of Nuclear Science and Engineering, Oregon State University, Corvallis, OR, 97331, USA\\
    }
}

\begin{document}

\maketitle

\begin{abstract}
    In Monte Carlo radiation transport calculations, Woodcock-delta tracking is a common alternative to the more popular surface tracking technique, where the largest cross section at a given energy in a problem is used to sample the distance to collision at any point.
    Because this process forces extra nonphysical collisions, it is paired with rejection sampling to determine real events from phantom events.
    Standard implementations of delta tracking preclude the use of a track- (or path-) length estimator for scalar flux tallies,  instead using the often higher-variant collision estimator.
    No mathematical reason prohibits use of the track-length estimator with delta tracking; however, algorithmic inefficiencies have made this combination rare. 
    In this work we introduce a delta-tracking algorithm that tallies fluxes to a structured rectilinear mesh using the track-length estimator.
    This development also enables hybrid surface-delta tracking algorithms, because the track-length tally can be used everywhere for scalar flux estimation regardless of which tracking algorithm is employed.
    We use this tallying technique to develop a novel hybrid-in-energy method, where delta tracking is used for high-energy particles (where mean free paths are long) and surface tracking is used for resonance energies and below. 
    We also implement a hybrid-in-material method, similar to what is implemented in Serpent2.
    We demonstrate that these delta tracking algorithms can be used in conjunction with continuously moving surfaces.
    We compare these methods showing figures of merit on four time-dependent problems (two multi-group and two continuous energy) solved with CPU- and GPU-based computers.
    Our implementation of delta tracking with a track length tally modestly improves figures of merit compared to standard delta tracking with a collision estimator and surface tracking with a track length estimator (\num{1.5}$\times$--\num{2.5}$\times$) for a problem with significant void regions.
    For both multi-group and continuous energy pressurized water reactor benchmarks, standard delta tracking with a collision estimator performs best.
    Hybrid-in-energy methods show significant improvements (\num{7}$\times$--\num{11}$\times$) for a continuous energy reactor benchmark problem.
\end{abstract}

\section{Introduction}

% general introduction
Predicting the neutron distribution in space, energy, and time is essential when modeling inertial confinement fusion experiments, pulsed neutron sources, and nuclear criticality safety experiments, among other systems.
The behavior of neutrons can be modeled with a Monte Carlo simulation, where particles are created and transported to produce a particle history \cite{Lewis1984}. 
The path of a particle and the specific set of events that occur within its history are governed by pseudorandom numbers, known probabilities (e.g., from material data), and known geometries. 
Data about how particles move and/or interact within the system are tallied to compute quantities of interest with an associated statistical error from the Monte Carlo process.

Two methods are commonly used to sample the random walk in a Monte Carlo neutron transport algorithm: surface tracking \cite{Lewis1984} and Woodcock delta tracking \cite{woodcock_1965_deltatracking}, which we simply call ``delta tracking'' for the remainder of this work.
These two tracking algorithms have complementary performance bottlenecks \cite{morgan_weighted-delta-tracking_2015}.
In geometrically complex models with many surfaces, surface tracking can require many expensive calculations to find the distance to a nearest surface which delta tracking avoids.
In contrast, when delta tracking in models with certain material compositions, the required rejection sampling can dominate computational cost.
Furthermore, standard implementations of delta tracking preclude evaluating quantities of interest with a track-length estimator, instead opting for a collision estimator, while surface tracking has no such restrictions \cite{leppanen_2017_collision}.
The track-length estimator often provides lower-variance estimates of quantities of interest (see Section~\ref{sec:tracking_algs_and_est}) \cite{mc2018}.
%Surface tracking requires potentially complicated geometric operations, whereas delta tracking does not.
Due to these differences, relative performance of these methods depend on the problem.
For a certain class of problems, a hybrid method may allow greater performance than either approach used individually.
There have been other recent developments on delta tracking---namely weighted delta tracking algorithms~\cite{molnar_variance_2018, morgan_weighted-delta-tracking_2015}---but in this work we explore variants of the non-weighted version.

% cite previous M&C paper

Previous research into hybrid delta-surface tracking on a structured mesh showed good performance for problems with complex arrangements of optically thin materials \cite{morgan_2021_mcatk}.
However, this work was limited to problems that define material regions and track on a structured mesh.
Making material-based decisions about when to use delta and surface tracking has also been explored and implemented in production Monte Carlo codes \cite{leppanen_development_2013, leppanen_2010_burnup, richards_monk_2015}.
Here, we extend the idea of \textit{tracking} on a structured mesh to full delta tracking (eliminating the distance-to-surface check almost entirely) and \textit{tallying} to a structured mesh, allowing the relatively efficient use of a track-length estimator for scalar flux.
We describe, verify, and evaluate the performance of a delta-tracking algorithm that allows the use of a track-length estimator on a structured tally mesh.
% MC/DC introduction

We then implement this delta tracking plus track length estimator combination in two hybrid delta-tracking schemes: one in which the choice of tracking algorithm is based on the material region, and another based on the particle energy.
We implement this work in Monte Carlo / Dynamic Code (MC/DC), an open-source Monte Carlo neutron transport application designed to conduct rapid numerical methods development, specifically for time-dependent problems \cite{morgan_monte_2024}.
We compute figures of merit for four computationally difficult time-dependent fixed source benchmark problems and run-time results for a whole CPU node (two Intel x86 Xeon Sapphire Rapids CPUs) and a whole GPU node (four Nvidia Tesla V100 GPUs) of supercomputing systems at Lawrence Livermore National Laboratory.

This work demonstrates the first published use of a track-length estimator for scalar flux with full delta tracking, the first use of delta tracking in conjunction with continuously moving surfaces, and the first use of a hybrid delta-tracking method based on particle energy.
The methods we present here will not be used to compute reaction rate tallies.
While this does limit the general applicability of these methods, scalar flux is a necessary quantity for hybrid methods (e.g., iterative quasi Monte Carlo \cite{Pasmann03062023, Pasmann04042025}) and other uses.

\section{Tracking Algorithms and Estimators}
\label{sec:tracking_algs_and_est}

% Surface tracking
Conducting Monte Carlo calculations in problems with heterogeneous materials requires some method of treating discontinuities in the cumulative probability distribution function~\cite{lux_1998}.
This is done with a tracking (or sampling) method. 
The first method we consider is surface tracking, described in Algorithm \ref{alg:si}. 
For a particle in material $m$, the distance to collision is sampled from a cumulative probability distribution function by
\begin{equation}
    d_{\text{collision}} = \frac{-\ln(\xi)}{\Sigma_{t,m}(E)} \;,
\end{equation}
where $\Sigma_{t,m}(E)$ [\SI{}{\per\centi\meter}] is the macroscopic total cross section at a given energy $E$ of the material $m$ and $\xi$ is a pseudorandom number between zero and one.

This sampling of the cumulative probability distribution function will only hold true while the particle is in material $m$.
If the distance to collision is beyond a material interface in a system with multiple materials, the particle must be stopped at that interface surface and a new distance to collision is sampled using the new material's $\Sigma_{t,m}(E)$.
This approach is an unbiased way to deal with the sampling of the distance to collision in a heterogeneous medium.
In a standard surface-tracking algorithm, both the distance to collision ($d_{\text{collision}}$) and the distance to the nearest surface along the particle's direction of travel ($d_{\text{surface}}$) are computed, and
the smaller of these two distances determines which event happens to the particle: a collision or a surface crossing.
After or while the particle is moving, tallies can be accumulated to compute quantities of interest.
If a collision occurs, more sampling and associated operations are performed (e.g., isotropic scattering of a particle off a particular constituent nuclide).
If the particle is still alive and has not exited the problem domain, the algorithm is repeated.
The computation of the distance to the nearest surface can become quite expensive as geometries grow in complexity (e.g., complex combinatorial solid geometries (CSG) or CAD-based surfaces).
Surface tracking is at the heart of many modern Monte Carlo neutron transport applications, including MCNP \cite{MCNP_RisingArmstrongEtAl}, Shift \cite{hamilton_continuous-energy_2019, pandya_implementation_2016, shift2}, GREAPMC \cite{gearpmc},  MONK/MCBEND \cite{richards_monk_2015}, MCATK \cite{mcatk}, Mercury \cite{POZULP2024},  McCARD \cite{mccard1, mccard2}, SCONE \cite{scone}, COG \cite{cog}, TRIPOLI-4/5 \cite{trip4, trip5}, and OpenMC \cite{romano_openmc_2015}.

\begin{algorithm}
\begin{algorithmic}[1]

    \State $m =$ look up material in current particle location

    \State $\Sigma_{t,m} =$ look up total macroscopic cross section of material $m$ 

    \While{particle is alive}
    
        \State $d_{\text{collision}} = -\ln{\xi} / \Sigma_{t,m} $
        
        \State $d_{\text{surface}} =$ compute distance to nearest surface along particle direction of travel

        \If {$d_{\text{collision}} < d_{\text{surface}}$}

                \State $d = d_{\text{collision}}$
                
                \State sample collision type
                
                \State carry out collision
                
             \Else 

                \State $d = d_{\text{surface}}$
        
                \State move particle to surface
    
                \State $m =$ lookup material on the other side of the surface
    
                \State $\Sigma_{t,m} =$ look up total macroscopic cross section of material $m$ 
    
                \If {surface is a boundary}
    
                    \State implement boundary condition
    
            \EndIf
        \EndIf
        \State score $d$ track lengths to tally bins
    \EndWhile
    
    \vspace{1.5em}
    \caption{A generic surface tracking algorithm.}
    \label{alg:si}
\end{algorithmic}
\end{algorithm}

Delta tracking is the next most common tracking approach, shown in Algorithm \ref{alg:trad}.
It starts by pre-processing a \textit{majorant} macroscopic cross section
\begin{equation}
    \maj(E) = \max\left({\Sigma_{t,1}(E), \Sigma_{t,2}(E), \dots, \Sigma_{t,m}(E), \dots, \Sigma_{t,M}(E)}\right) \; ,
\end{equation}
where $M$ is the total number of materials in the simulation, such that the majorant is the largest cross section in any material in the problem.
Cross section data libraries for surface tracking codes often have separate energy grids specialized to each individual nuclide.
To compute the correct macroscopic majorant cross section we use a two-step interpolation method.
First, we compute the macroscopic total cross section of each nuclide in a given material, then unify the energy grid and interpolate all cross section points onto that grid.
This is repeated when going from single materials to an energy grid for all materials within a problem.
The algorithm implemented in MC/DC is shown in Appendix~\ref{app:majorant}.
Figure~\ref{fig:majorant_c5ce} shows a macroscopic majorant cross section for a typical pressurized water reactor as a function of neutron energy.
When using delta tracking, the distance to collision is always sampled with the majorant,
\begin{equation}
    d_{\text{collision}} = \frac{-\ln{(\xi)}}{\maj{}(E)} \; .
\end{equation}
However, this sampling forces extra collisions that do not physically occur; using the majorant generates the smallest distance to collision.
Delta tracking algorithms use a rejection sample to determine whether the collision event is real or a ``virtual'' (sometimes called ``phantom'') collision.
At the current location of the particle, the true material is identified to compute $\Sigma_{t,m}$ after a potential collision.
If 
\begin{equation}
    \xi > \frac{\Sigma_{t,m}(E)}{\maj(E)} \; ,
\end{equation}
where $\xi$ is a new random number, the collision did not physically occur and is rejected, so the particle can be left uncollided, alive with its current direction of travel, weight, and energy.
From here, the process is the same as before: tallies are accumulated, collision physics is carried out when appropriate, and the algorithm continues as long as the particle is still alive.

\begin{figure}
    \centering
    \includegraphics[width=0.9\textwidth]{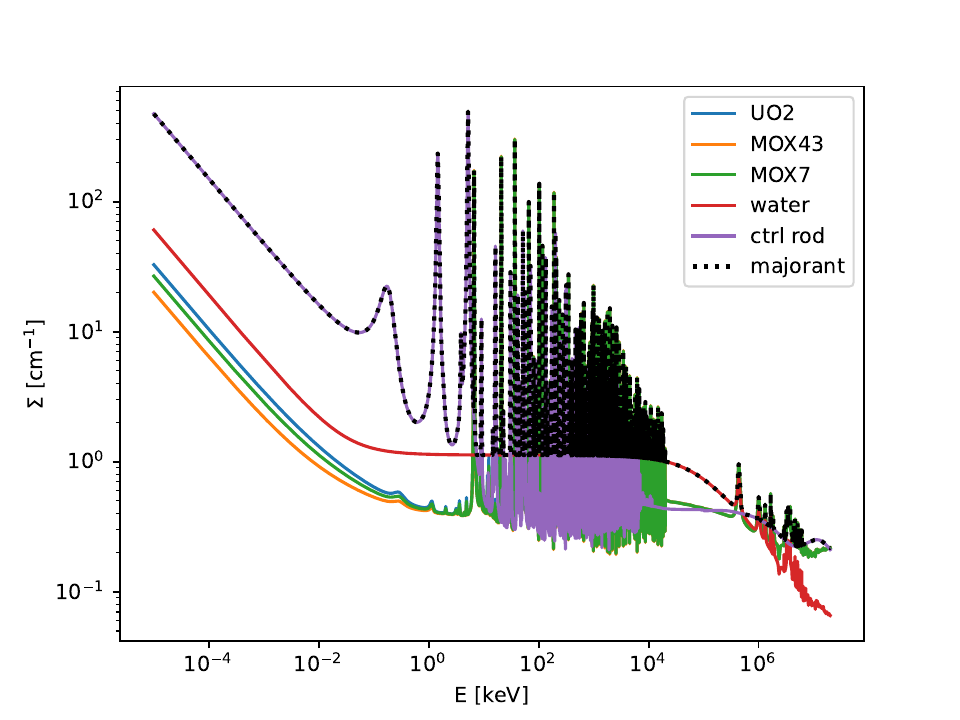}
    \caption{Continuous energy macroscopic majorant cross section $\maj(E)$ and other material cross sections for the continuous energy version of C5G7 pressurized water reactor described in Appendix \ref{app:c5ce_mat}.}
    \label{fig:majorant_c5ce}
\end{figure}

Delta tracking requires some way to kill or reflect particles on vacuum or reflecting boundaries, respectively.
In a code that already implements surface tracking, the functionality to track distances to specific surfaces will already exist.
So, it is natural when utilizing delta tracking to exclusively compute and assess distances to boundary surfaces.
If the number of boundary surfaces is small and/or the surfaces can be represented by simple functions, this calculation is computationally trivial.
%This is similar to other methods currently under development in OpenMC.

\begin{algorithm}
\begin{algorithmic}[1]

    \While{particle is alive}
    
        \State $d_{\text{collision}} = -\ln{\xi} / \maj $
        
        \State $d_{\text{boundary}} =$ compute distance to boundary

        \If {$d_{\text{collision}} < d_{\text{boundary}}$}

            \State move particle to collision site

            \State $m =$ look up material in current particle location
            
            \State $\Sigma_{t} =$ look up total macroscopic cross section of material $m$ 

            \If { $\xi > \Sigma_{t} / \maj$ }

                \State collision is rejected

             \Else 
            
                \State collision is accepted

                \State tally $1/\Sigma_t$ to bin at particle's current location

                \State determine collision type 
                
                \State carry out collision
                
            \EndIf
                
        \Else 
        
            \State move particle to boundary

            \State implement boundary condition
        \EndIf
        
    \EndWhile
    
    \vspace{1.5em}
    \caption{Delta tracking in MC/DC. Notably, we are still surface tracking to boundaries.}
    \label{alg:trad}
\end{algorithmic}
\end{algorithm}

An added complication when using delta tracking is that only certain tallies can be efficiently scored.
Many estimators can be used to compute quantities of interest (e.g., scalar flux, reaction rates) by tallying events that occur within a given region of phase space.
Two common scalar flux estimators are the collision estimator and the track-length (also known as the path length), estimator \cite{Lewis1984}.
Often, tallies are scored into bins on a structured mesh that overlays the surfaces and material regions that a Monte Carlo simulation uses to conduct actual transport operations.
The track-length estimator is
\begin{equation}
    \label{eq:pathlength}
    \hat{\phi}_n = \sum_{i=1}^{I}p_{i,n} \; ,
\end{equation}
where $\hat{\phi}$ is the scalar flux integrated over a given mesh cell $n$, $p_{i,n}$ is the track length of particle $i$ passing through mesh cell $n$, and $I$ is the total number of particles.
The collision estimator is
\begin{equation}
    \label{eq:collision}
    \hat{\phi}_n = \sum_{i=1}^{I} \frac{1}{\Sigma_{t,n}(E)} \;.
\end{equation}
The collision estimator will often produce a flux estimate with higher variance compared to the track-length estimator when $\Sigma_{t,m}$ is small (i.e., optically thin, less-dense materials) and is characterized by a zero tally in a true void region \cite{mc2018, leppanen_2017_collision}.
On the other hand, the track-length estimator will tally into every mesh bin the particle moves through during its random walk.
In some problem regimes, more information will be scored for a given number of particle histories, which will result in a tally with lower variance.

The estimators in Eqs.~\eqref{eq:pathlength} and ~\eqref{eq:collision} only apply to flux integrals.
Response functions for other reaction rates (e.g., fission rate density) will require knowledge of the macroscopic cross section of that reaction in a given location \cite{lux_1998}.
In this work, when using the track-length estimator with delta tracking we limit ourselves to flux integrals only.
%So, we also only enable these schemes for fixed-source problems and leave $k$-eigenvalue calculations for future work.
%This does currently limit the more general applicability of the methods we implement, though future work may extend these methods to allow computing of these parameters.
We discuss this further in Section~\ref{disucssions}.

Delta-tracking algorithms are implemented in many production Monte Carlo neutron transport applications, including Serpent2 \cite{leppanen_2010_burnup, leppanen_2017_collision, leppanen_development_2013, leppanen_2015_serpent}, MONK/MCBEND \cite{richards_monk_2015}, SCONE \cite{scone}, IMPC \cite{fang_development_2022}, TRIPOLI-5 \cite{trip5}, and GUARDYAN \cite{molnar_guardyan_2019}.
Notably, the MONK Monte Carlo neutron transport code is the direct successor to the GEM code where Woodcock et al.\ first implemented delta tracking \cite{woodcock_1965_deltatracking}.

% what other codes do and how this work is novel
Modern transport applications often involve only one of these two tracking algorithms, and the implementation of their code base is optimized from there.
However, either method of sampling the probability distribution functions is valid for the same system in the same simulation, even at the same location in phase space.
Some developers have taken advantage of this logic.
For example, Serpent2 Monte Carlo code establishes regions of delta tracking and surface tracking based on the ratio between $\Sigma_{t,m}$ and $\maj$ and a user-supplied constant $\in [0,1]$ (where $0$ means delta tracking only and $1$ surface tracking only) \cite{leppanen_development_2013}.
MONK/MCBEND supports material regions inside of which delta tracking is implemented (called hole geometries) for complex materials like stochastically heterogeneous materials (e.g., TRISO fuel elements, aggregate and cement mixtures in concrete) while surface tracking is used elsewhere \cite{richards_monk_2015, richards_monk_2025}.
%Ongoing developments in OpenMC and MCNP will introduce similar delta tracking algorithms for 
%Publications indicate that no other Monte Carlo neutron transport application currently supports the use of any track length estimator within a region undergoing delta tracking.

\section{Hybrid Delta Tracking Schemes}

In this section, we introduce and verify the voxelized tally structure that allows the implementation in MC/DC of a track-length estimator with delta tracking.
We also introduce two hybrid surface-delta tracking schemes that we implemented for time-dependent transport with moving surfaces in MC/DC.
The first is a region-based delta tracking method we call ``hybrid-in-material'' that has been previously explored in other production Monte Carlo neutron transport applications \cite{fang_development_2022, leppanen_development_2013}.
The second method uses delta tracking only for neutrons with energies above resolved neutron cross section resonances, where total cross sections are often similar to the majorant, and mean free paths are long relative to nominal system dimensions.
This approach is called ``hybrid-in-energy''.

\subsection{Voxelized Tallies}

Standard delta-tracking algorithms do not require information about the material or physical mesh cell a given particle occupies at any moment in transport.
Conventional wisdom dictates that repeatedly determining locations, cross sections, and tally bin indices while using delta tracking would be costly and eliminate any benefit to performance that the method provides \cite{mc2018}.
%It follows that delta tracking precludes the use of a track-length estimator for quantities of interest.
However, there is nothing \textit{mathematically} preventing the use of a track-length estimator with delta tracking.
Furthermore, if all that is desired as a quantity of interest for a given problem is the scalar flux, no cross section lookups are needed while tallying---only the physical track length and corresponding bin.
This forms the underlying idea behind the voxelized tally method.
We investigate whether efficient implementation of tallying allows for effective use of a track length estimator when delta tracking and if this results in a performance benefit.

\begin{figure}[htbp]
  \centering
  \includegraphics[width=\textwidth]{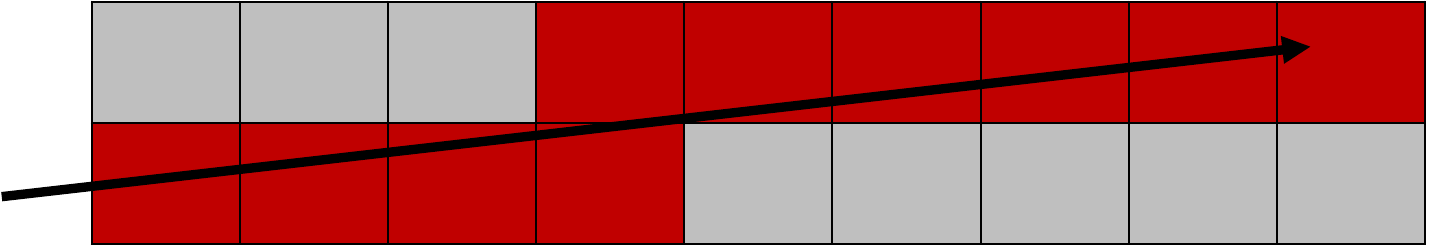}
  \caption{A particle tallying to multiple tally mesh bins (shown in red). Implemented as a single operation for both surface and delta tracking in MC/DC.}
  \label{fig:tally_ray}
\end{figure}

As of v0.11.1, MC/DC \cite{morgan_monte_2024,transport_cement_mcdc_2024} has a tally algorithm that scores track lengths crossing multiple structured mesh tally bins (voxels) in a single operation.
This algorithm is based on a sweep method in which the initial state [mesh cell index, ($x$, $y$, $z$) position, direction of travel, speed, and particle clock] is known, as is the distance to the next event.
The particle is then swept from tally voxel to tally voxel, accumulating the track length traveled in a given voxel along the way.
Figure~\ref{fig:tally_ray} shows a hypothetical particle track and the voxels to be tallied (in red) in a single operation.
In MC/DC's surface-tracking algorithm, this occurs even before moving a particle to an event.
This is computationally efficient as we are tallying to a voxel---a structured rectilinear mesh bin.

In this work, we use this voxelized tally scheme when undergoing delta tracking.
Even if a particle collision is rejected, that particle still physically moves to the location that was sampled, allowing the use of the track-length estimator.
%In this scheme, the distance tallied is always the distance sampled with the majorant, aside from census crossings or distance-to-boundary events.
We expect this voxelized tally method to produces scalar fluxes with smaller variance in problems with optically thin regions, where a collision estimator (adapted or otherwise) will not perform efficiently.

\begin{figure}[htbp]
  \centering{
  \begin{subfigure}[b]{0.49\textwidth}
      \includegraphics[width=\textwidth]{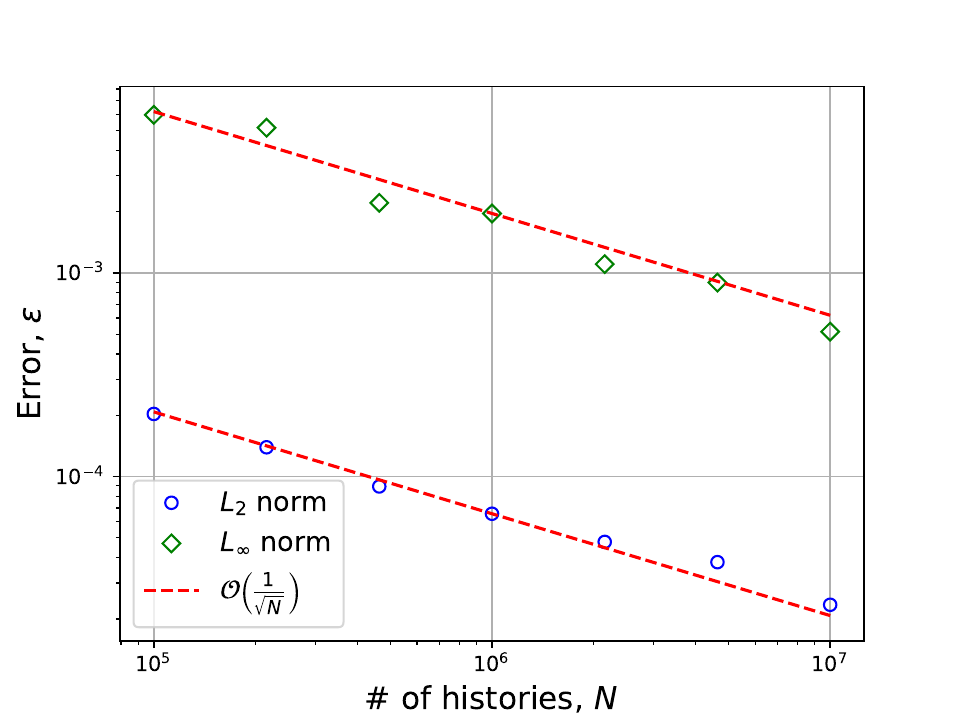}
      \caption{Critical problem with $\epsilon$ from scalar flux.}
  \end{subfigure}
  \hfill
  \begin{subfigure}[b]{0.49\textwidth}
      \includegraphics[width=\textwidth]{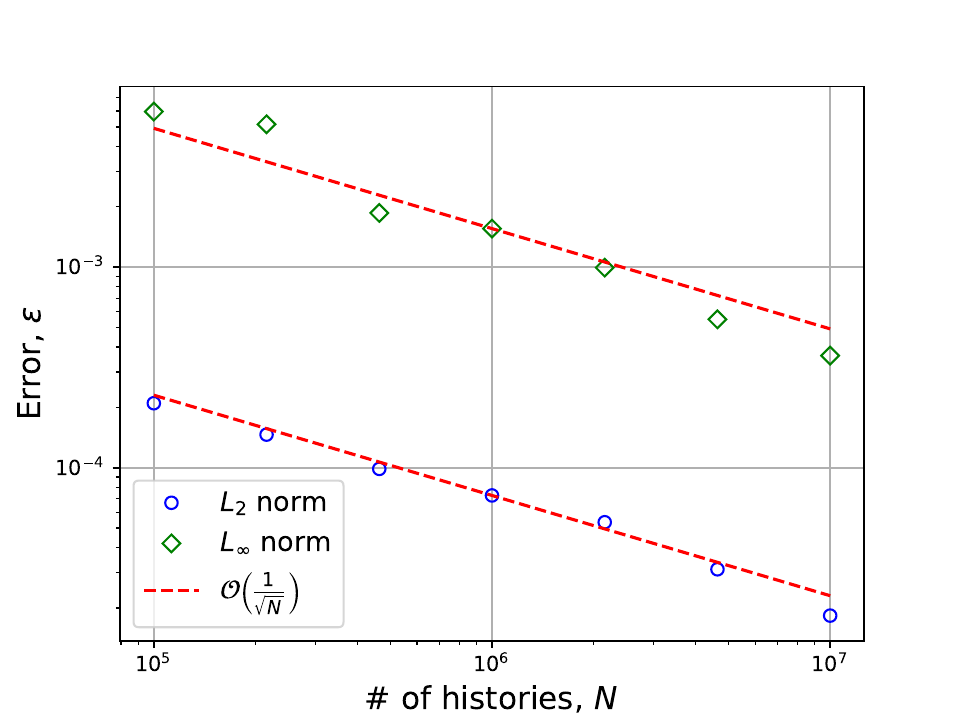}
      \caption{Critical problem with $\epsilon$ from scalar flux at census.}
  \end{subfigure}
  \hfill
  \begin{subfigure}[b]{0.49\textwidth}
    \includegraphics[width=\textwidth]{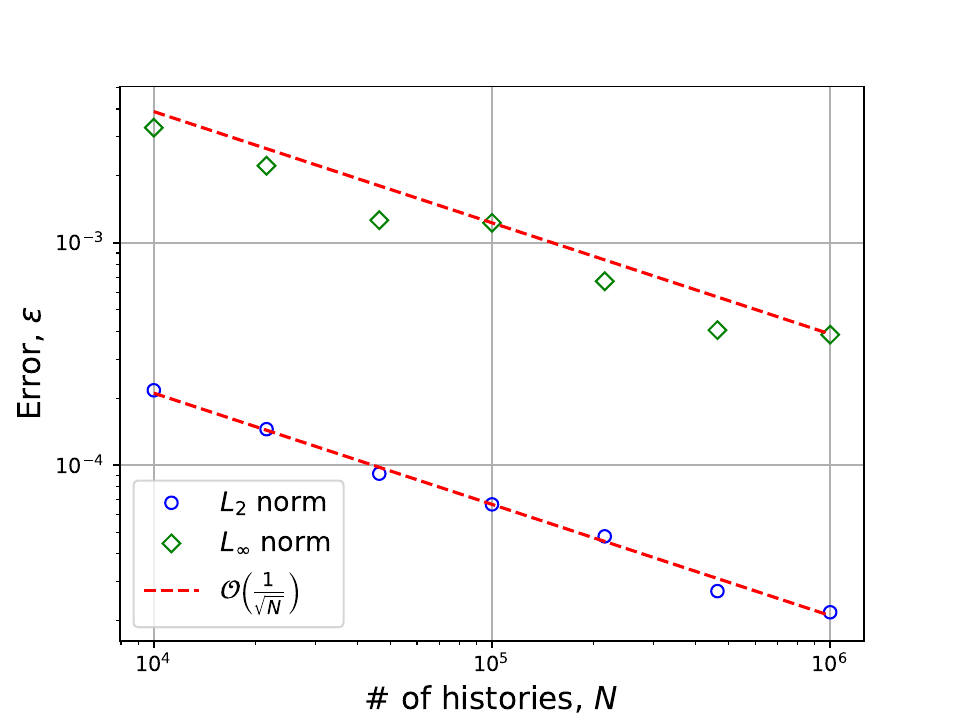}
    \caption{Subcritical problem with $\epsilon$ from scalar flux tallies at census.}
  \end{subfigure}
  \hfill
  \begin{subfigure}[b]{0.49\textwidth}
    \includegraphics[width=\textwidth]{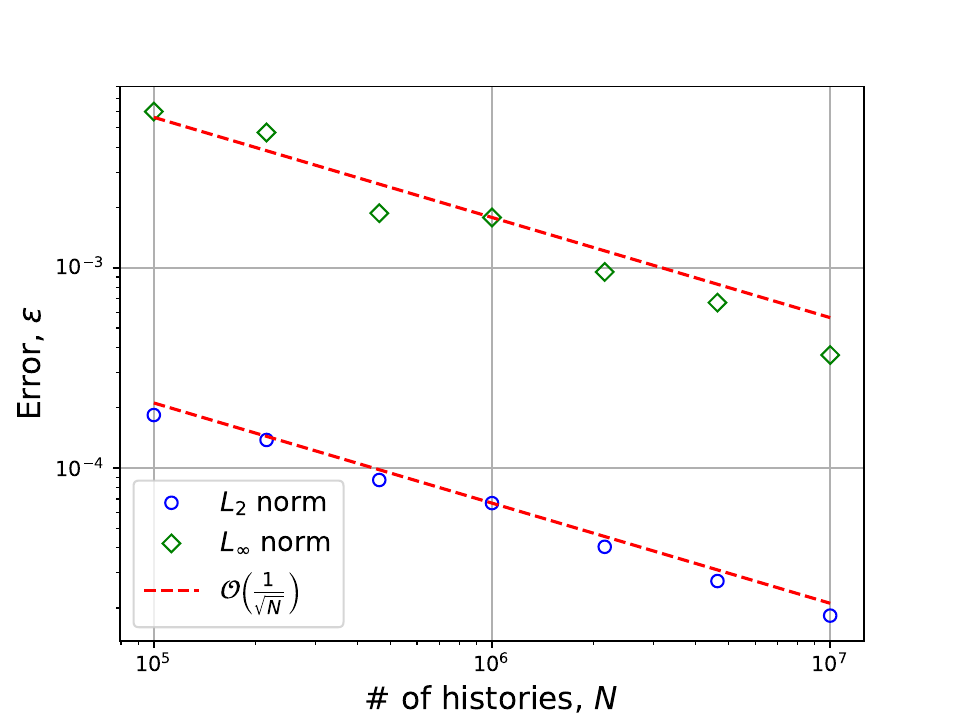}
    \caption{Subcritical problem with $\epsilon$ from scalar flux. \\ \hfill}
  \end{subfigure}
  \hfill
  \begin{subfigure}[b]{0.49\textwidth}
    \includegraphics[width=\textwidth]{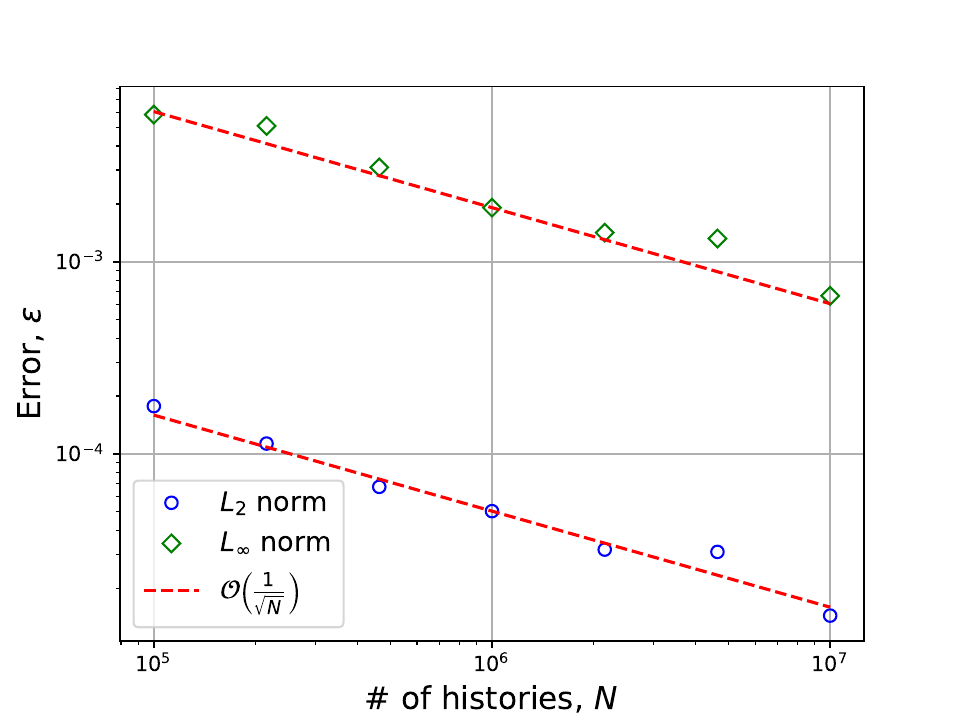}
    \caption{Supercritical problem with $\epsilon$ from scalar flux.}
  \end{subfigure}
  %\hfill
    \caption{Convergence rate verification of AZURV1 \cite{ganapol_2001_homogeneous}, showing the expect Monte Carlo convergence rate ($N^{-1/2}$).}
  \label{fig:azurv1}}
\end{figure}

\begin{figure}[htbp]
  \centering
  \begin{subfigure}[b]{0.49\textwidth}
  \includegraphics[width=\textwidth]{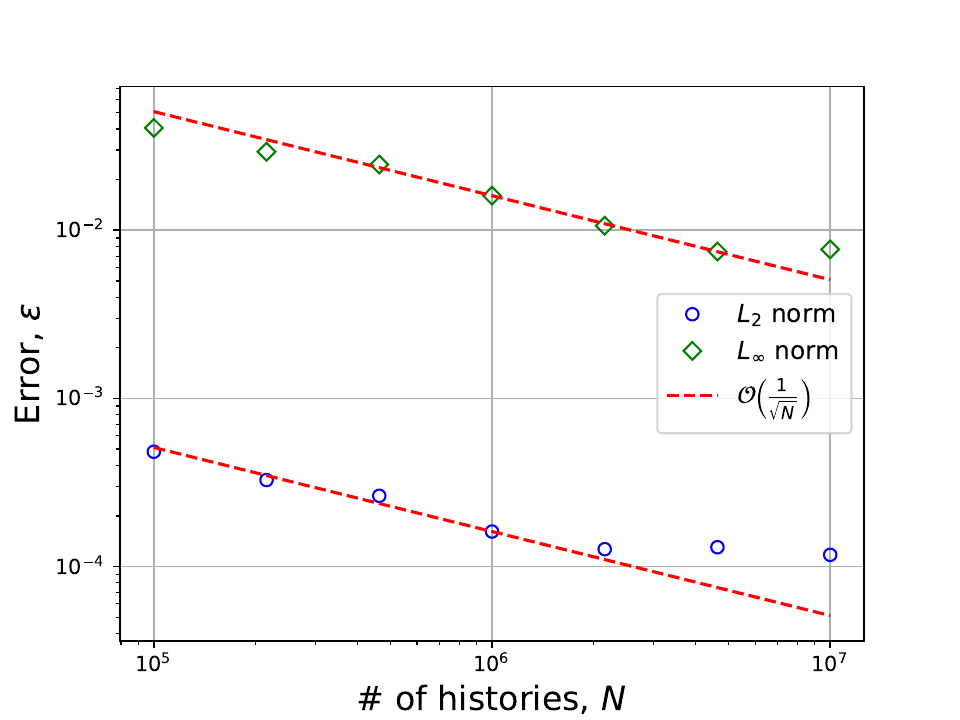}
  \caption{Steady-state problem with $\epsilon$ from scalar flux.}
  \end{subfigure}
  \hfill
  \begin{subfigure}[b]{0.49\textwidth}
  \includegraphics[width=\textwidth]{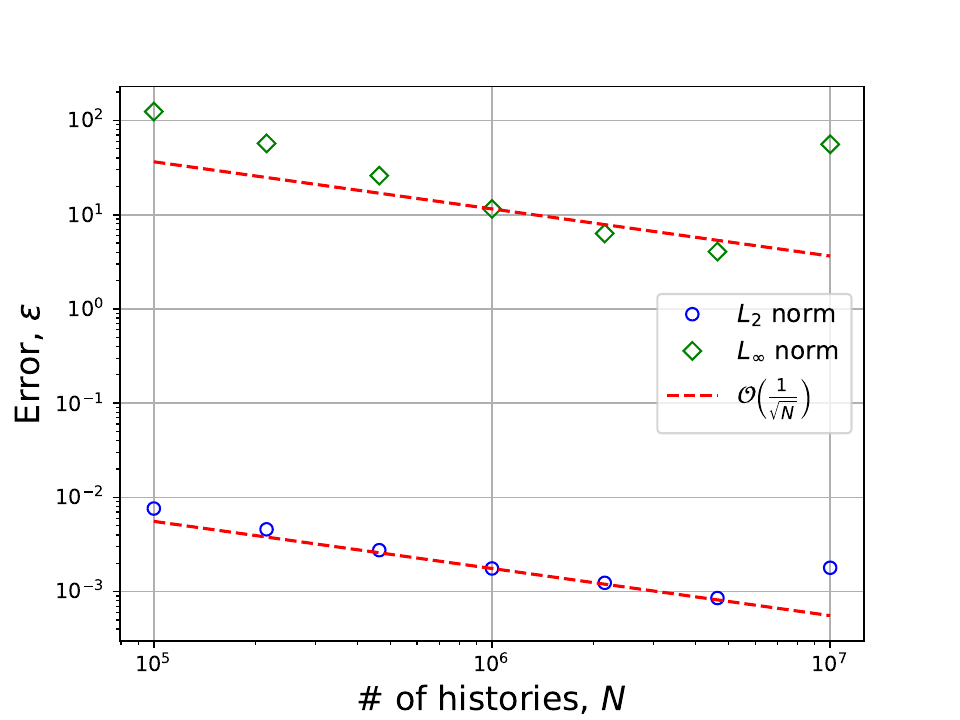}
  \caption{Time-dependent problem with $\epsilon$ from scalar flux.}
  \end{subfigure}
  \hfill
  \begin{subfigure}[b]{0.49\textwidth}
  \includegraphics[width=\textwidth]{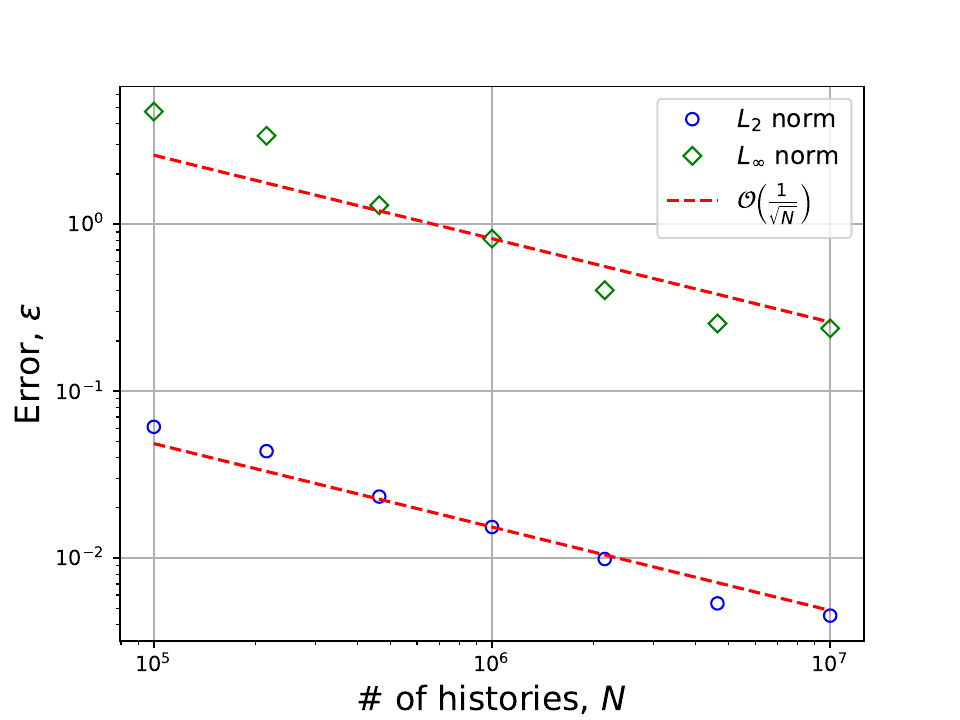}
  \caption{Time-dependent problem with $\epsilon$ from neutron density.}
  \end{subfigure}
  \hfill
  \begin{subfigure}[b]{0.49\textwidth}
  \includegraphics[width=\textwidth]{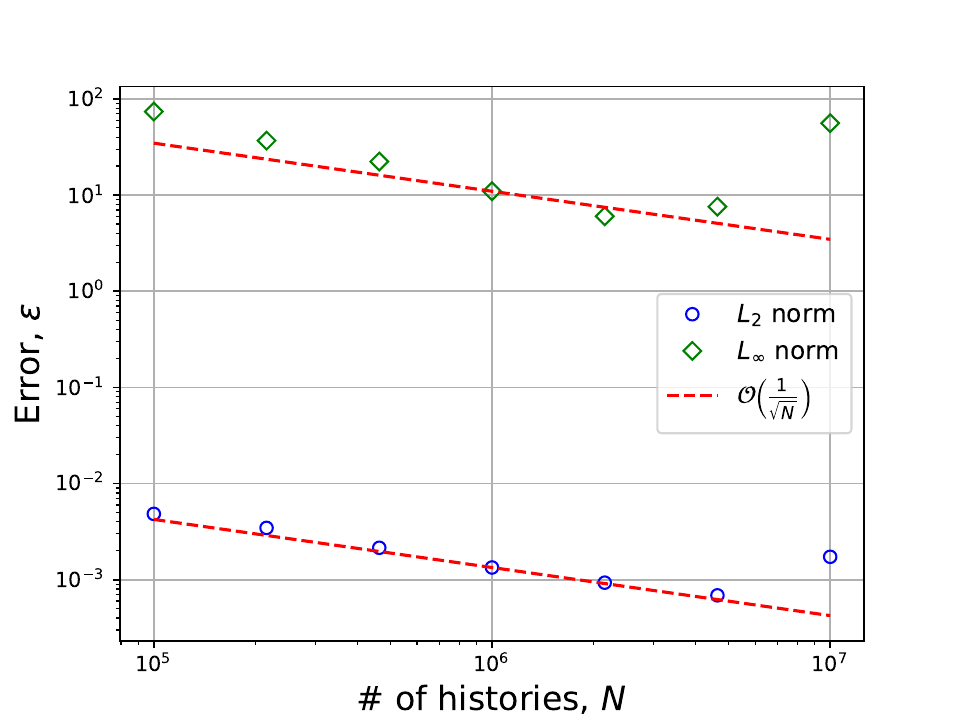}
  \caption{Time-dependent problem with $\epsilon$ from flux at census.}
  \end{subfigure}
  \hfill
  \begin{subfigure}[b]{0.49\textwidth}
  \includegraphics[width=\textwidth]{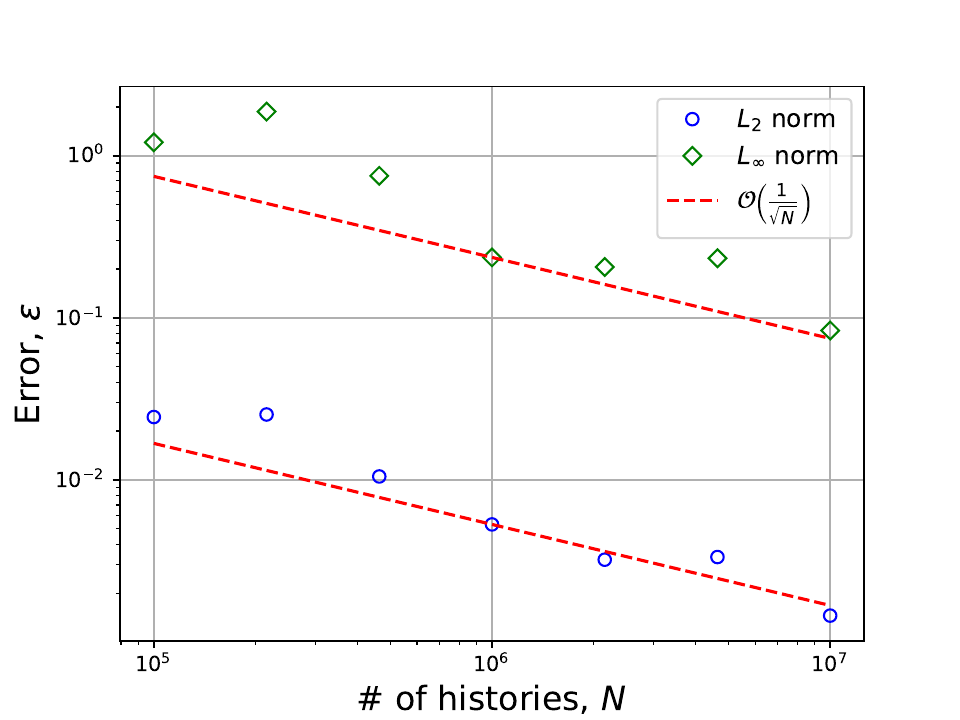}
  \caption{Time-dependent problem with $\epsilon$ from neutron density at census.}
  \end{subfigure}
  \caption{Convergence rate of an infinite pin using the SHEM 361 group cross section library \cite{hfaiedh_2005_shem}, showing the expected Monte Carlo convergence rate ($N^{-1/2}$).}
  \label{fig:shem}
\end{figure}

\begin{figure}[htbp]
  \centering
  \includegraphics[scale=0.75]{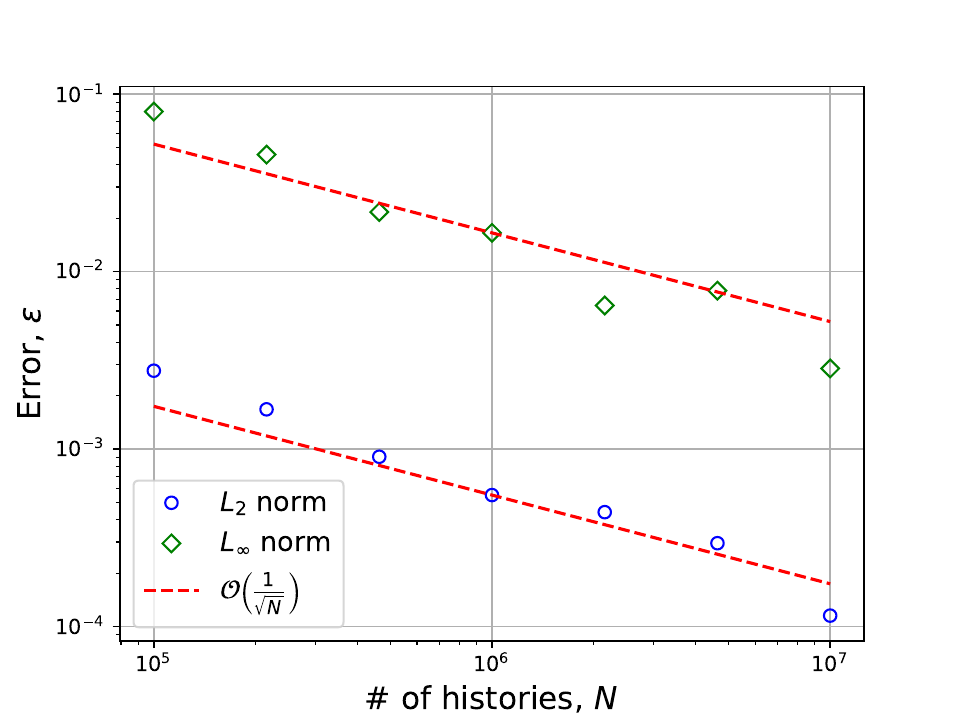}
  \caption{Convergence rate of flux from Reed's problem \cite{reed_difference_1971} showing the expect Monte Carlo convergence rate ($N^{-1/2}$).}
  \label{fig:reeds}
\end{figure}

\begin{figure}[htbp]
    \centering
    \includegraphics[width=0.75\linewidth]{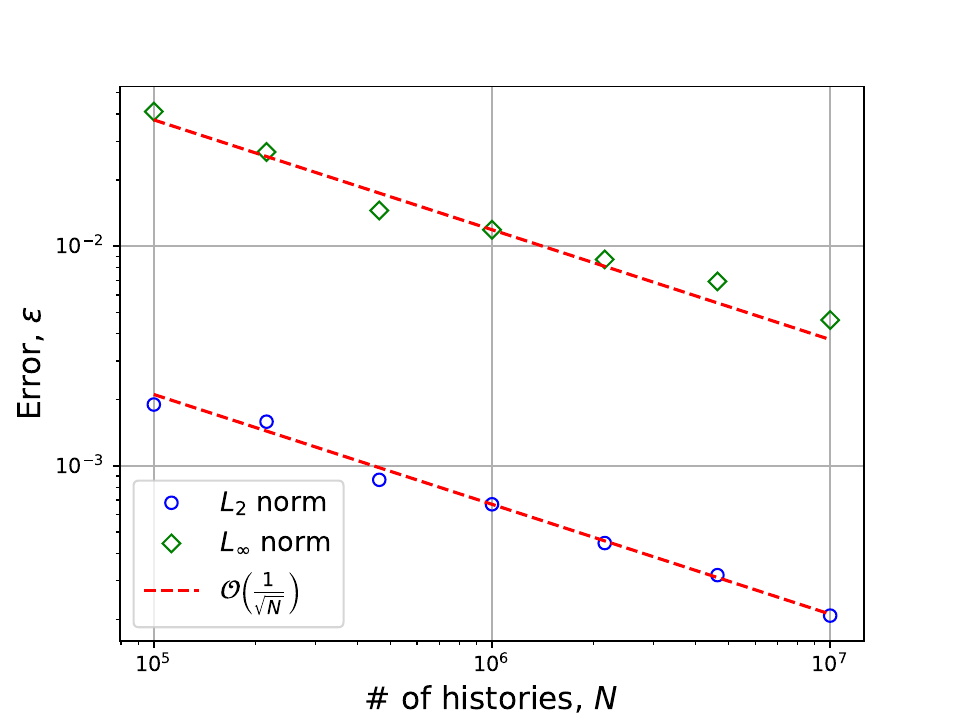}
    \caption{Convergence rate of quantities of interest for a purely absorbing slab wall problem, showing the expect Monte Carlo convergence rate ($N^{-1/2}$).}
    \label{fig:abs_slab}
\end{figure}

To verify that MC/DC's voxelized track-length tallies with delta tracking converge to the correct solution and at the correct rate, we use four analytic fixed-source benchmark problems from MC/DC's verification suite.
A fixed-source calculation is well posed when the location of the source is known and an initial condition is proved (i.e. not $\alpha$ or $k$ eigenvalue calculations).
We compare the error ($\epsilon$) in both the $L_\infty$ and $L_2$ norms, from integral quantities of interest to a reference solution and plot the error as a function of increasing particle count.
The convergence rate should be the standard Monte Carlo convergence rate of $N^{-1/2}$.
Our verification problems are:
\begin{itemize}
    \item AZURV1 time-dependent benchmark both super and sub critical (Figure~\ref{fig:azurv1}) \cite{ganapol_2001_homogeneous};
    \item A time-dependent infinite pin cell using the 371 group SHEM cross sections (Figure~\ref{fig:shem}) \cite{hfaiedh_2005_shem}; 
    \item Reed's problem (Figure~\ref{fig:reeds}) \cite{reed_difference_1971}; and
    \item A purely absorbing three-region slab (Figure~\ref{fig:abs_slab}).
\end{itemize}
All verification simulations show the $N^{-1/2}$ convergence rate expected for Monte Carlo results.
This verifies that we get the expected results with voxelized delta tracking.

\subsection{Hybrid-in-Material}
\label{sec:material_exc}

The first hybrid method we implement in MC/DC is the hybrid-in-material method.
Each active particle is assigned an additional flag that declares which transport algorithm it is using to sample a distance from the next event.
%At the beginning of the \texttt{determine\_next\_event()} function, the tracking flag will be set to \texttt{true} if the particle is in a material region declared by the user to undergo delta tracking or \texttt{false} if in a region where delta tracking should not be used. 
This is similar to Serpent2's algorithm \cite{leppanen_2017_collision}; however, our voxelized tally structure allows us to use track-length estimators in all regions, not just those in which the particle is undergoing surface tracking.
Also, Serpent2's algorithm makes automatic decisions about where to surface and delta track based on a single user-supplied cut-off value between zero and one for the whole problem \cite{leppanen_development_2013}, whereas we leave it the user to specify on a material-by-material basis.
A more automatic approach similar to Serpent2 will be implemented in future work.
The IMPC Monte Carlo neutron transport code also implements a similar hybrid-in-material surface-delta tracking scheme \cite{fang_development_2022}.

\begin{figure}[htbp]
    \centering
    \includegraphics[width=.95\textwidth]{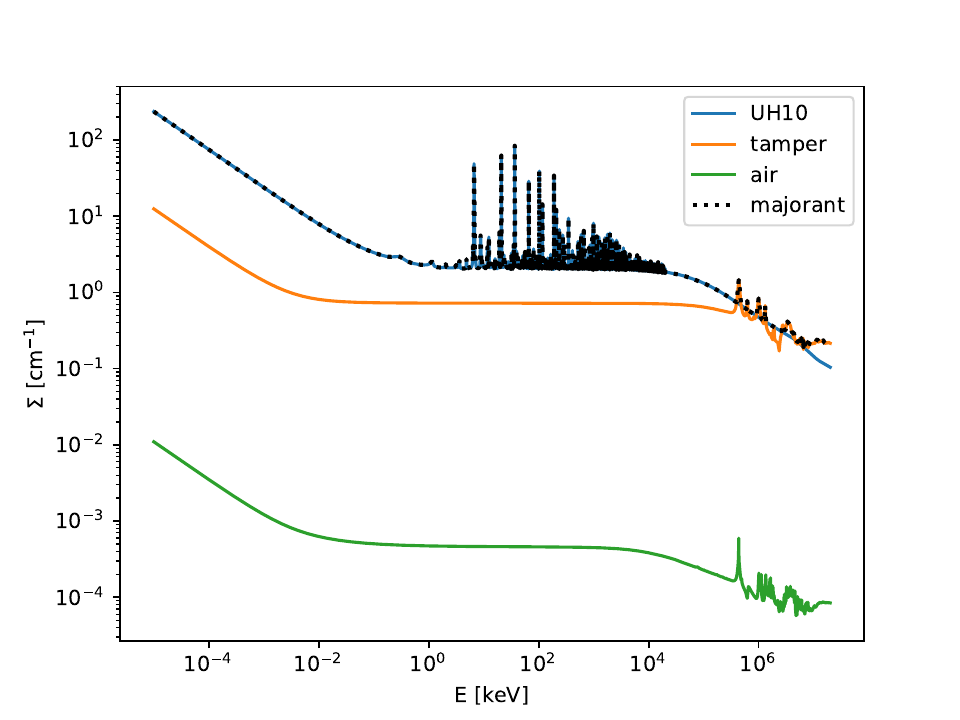}
    \caption{Continuous energy macroscopic majorant ($\maj$) and other material cross sections for the Dragon burst problem.}
    \label{fig:majorant_dragon}
\end{figure}

%Conventional wisdom dictates that delta tracking should not be used in systems with strong localized absorbers as the majorant will be governed by cross sections much larger than others in the system.
Delta tracking is typically inefficient in systems with strong localized absorbers because particles experience a large number of virtual collisions.
For example, Figure~\ref{fig:majorant_dragon} shows the cross section data for the Dragon burst problem we describe in Section~\ref{sec:benchmarks}.
This is a simulation that does not warrant delta tracking.
This simulation contains three materials, two of which (the fuel and tamper) have similar cross sections over all energies.
However, the cross sections for air are over four orders of magnitude lower than the cross sections for the fuel and tamper materials.
This means that while particles are delta tracking in the air, they will get stuck in the rejection-sampling loop, statistically rarely completing a particle history.
Furthermore, for standard delta tracking (which uses only a collision estimator), the tallies in the air will have higher variance as there will be statistically fewer collisions taking place.
Using surface tracking in the air and delta tracking in the material may improve performance in this problem.

\subsection{Hybrid-in-Energy}
\label{sec:cutoff}

Most high-energy neutron interactions with materials are characterized by relatively long mean free paths (i.e., small cross sections) because they primarily undergo potential scattering \cite{duderstadt_hamilton}.
In systems with a large number of surfaces and fast neutrons, delta tracking generally does better compared to surface tracking, because surface tracking gets stuck moving particles from region to region.
Delta tracking, on the other hand, will stream particles through the whole problem, never conducting expensive distance-to-nearest-surface computations.
Furthermore, at higher energies, cross sections tend to vary less across material interfaces, alleviating issues with the rejection-sampling loop.  These observations motivate the development of the ``hybrid-in-energy'' method that employs delta tracking for high energy neutrons and surface tracking, otherwise.

For example, consider a continuous energy version of the C5G7 benchmark reactor geometry \cite{hou2017oecd}.
Due to the lack of isotopes available in MC/DC's cross section library To simplify the problem to allow for rapid numerical methods exploration we replace the guide tube and fission chamber materials with moderator material. We also replace MOX fuels with UO$_2$ with different enrichment.
The material composition is given by Table~\ref{tab:c5ce} in Appendix~\ref{app:c5ce_mat}.
Figure~\ref{fig:majorant_c5ce} shows the macroscopic cross sections for the seven-material reactor, with the majorant in black.
Around \SI{10}{\kilo\electronvolt} the neutron resonances end and the system is ``high energy''. 

If using delta tracking for this whole problem, the increased absorption in the resonances of some materials exacerbates issues with the required rejection sampling. 
Therefore, for this problem, it would be ideal to delta track above \SI{50}{\kilo\electronvolt} (to completely avoid neutron resonances) and surface track under that threshold.
This forms the basis for our hybrid-in-energy method where both surface and delta tracking are used depending on the energy of individual particles.

\subsection{Implementation in MC/DC}
\label{sec:implementation}

MC/DC~\cite{morgan_monte_2024} is designed to implement and test new time-dependent Monte Carlo numerical methods at scale \cite{variansyah_mc23_mcdc}.
It uses a novel (for the field) development structure in which Python compute kernels are compiled via the Numba just-in-time compiler to run on CPUs and with the Harmonize runtime manager \cite{brax2023, morgan_2025_monte, cuneo_2025_harmonize} to run on GPUs.
MC/DC enables rapid experimentation with these methods at scale on both CPUs and GPUs with time-dependent problems of interest.

Generally, to implement delta tracking in a code that already implements surface tracking, the operations needed are:
\begin{enumerate}
    \item Pre-process functions to generate a majorant (MC/DC's implementation is described in Appendix~\ref{app:majorant});
    \item Add \texttt{if} statements in \texttt{distance\_to\_next\_event()} functions to compute relevant distances;
    \item Add functions to implement boundary conditions when in delta-tracking mode; and
    \item Elevate delta-tracking options to the input deck.
\end{enumerate}
This process is similar to that used to implement hybrid surface-delta tracking methods in MCATK~\cite{morgan_2021_mcatk}.
However, previous work in MCATK implemented a hybrid delta tracking algorithm, which was conceived in part to integrate into MCATK smoothly.

In MC/DC, we have implemented standard delta tracking, delta tracking using voxelized tallies, and two hybrid delta-surface tracking algorithms.
Our implementation in total added about \num{450} lines of code, of which half involved computing various types of majorants (an example can be found in Appendix~\ref{app:majorant}). 
Only about \num{200} lines of code were required in the compute kernels to implement all considered methods.

%This is also similar to current ongoing work in OpenMC.

When converting a surface tracking code to enable delta tracking, a simple distance-to-nearest boundary function can be produced using existing functionality to find the distance to a nearest surface.
In effect, we are still surface tracking---but only to the reflecting boundary surfaces.

%We found that the most complicated issue when implementing standard Woodcock delta tracking in MC/DC was handling boundary conditions.
%Surface tracking codes (like MC/DC) often contain boundary condition flags for all surfaces which will be flipped to \texttt{True} for boundary surfaces.
%When Woodcock delta tracking we will not be able to use this information.

%Vacuum boundary conditions are simple;  if a particle is determined to be outside of the problem domain when looking for a cross section during the rejection sample, particles are terminated.
%For reflecting surfaces we use a stripped down version of the \texttt{distance\_to\_nearest\_surface} function in MCDC to compute distances to reflecting surfaces only, if any exist in a given problem.
%Many reflecting boundary conditions, including those that exist in our benchmark problems, are imposed on planar surfaces, making these specific distance to boundary computations computationally inexpensive.
%In effect, we are still surface tracking---but only to the reflecting boundary surfaces.

\section{Verification of delta tracking with continuously moving surfaces}

\begin{figure}
    \centering
    \includegraphics[width=.9\linewidth]{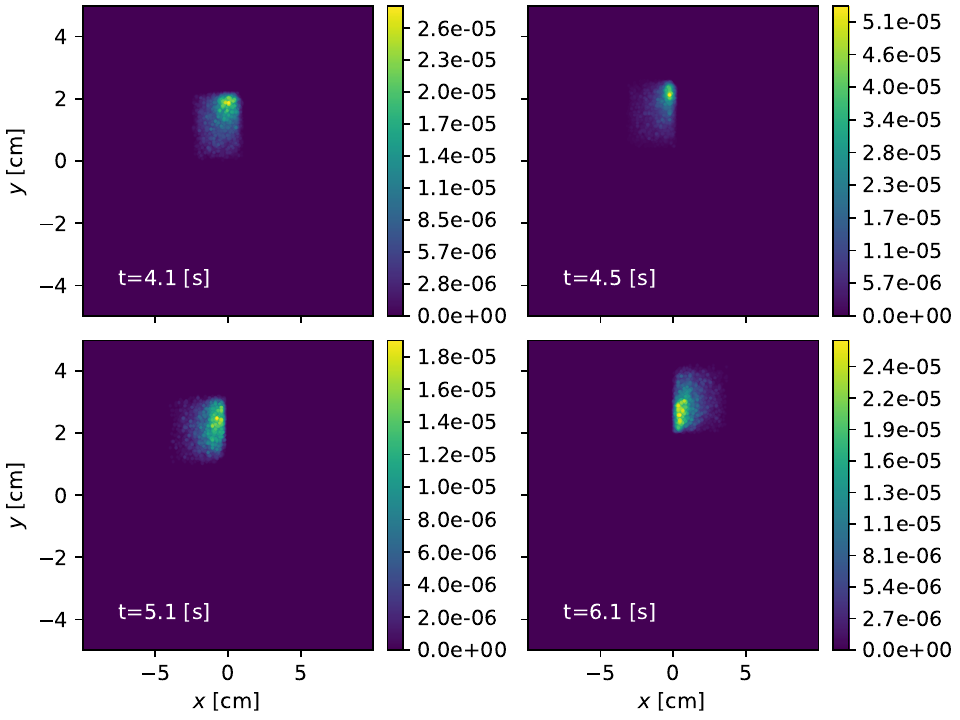}
    \includegraphics[width=.85\linewidth]{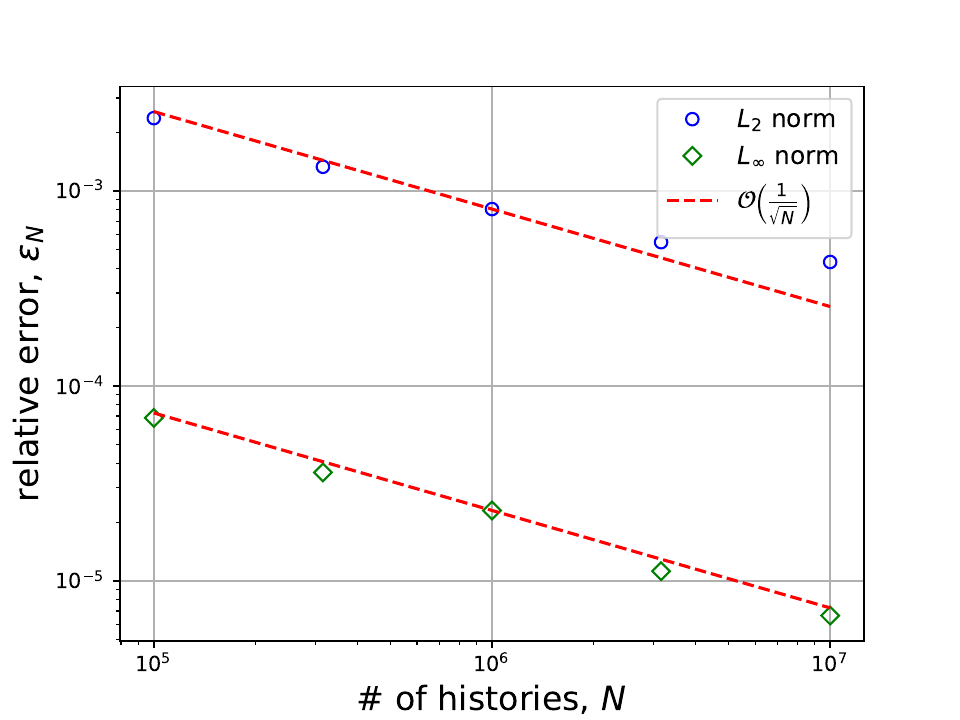}
    \caption{For the moving pellet problem (top) fission reaction rate density at various points in time, (bottom) convergence between fluxes produced from surface and delta tracking both with the track-length estimator, showing $N^{-1/2}$ convergence rate.}
    \label{fig:moving_pellet}
\end{figure}

To verify that delta tracking can be used in conjunction with continuously moving surfaces in MC/DC, we use the moving pellet test from MC/DC's regression test suite \cite{morgan_monte_2024}: a cylindrical fuel element moves through a region that also has a small source.
As the pellet moves closer to, then further away from, the source region, the fission rate in the pellet changes.
Figure~\ref{fig:moving_pellet} at top shows the fission reaction rate density at various points in time.
The outline of the pellet is clearly visible in a number of time steps.
These plots were produced using delta tracking with a collision estimator (to compute fission reaction rates).

To further confirm that delta tracking may be used with continuous movement physics, we compare the scalar flux solutions provided from surface tracking and delta tracking with voxelized tallies at various particle counts.
We compute 
\begin{equation}
    \epsilon_N = \left|\left|\phi_N^{\text{surface}} - \phi_N^{\text{delta}} \right|\right|_2 \;,
\end{equation}
where $\phi_N^{\text{surface}}$ and $\phi_N^{\text{delta}}$ are scalar flux computed by surface and delta tracking, respectively, at every choice of $N$ particles.
We then compare the norm of the two results over particles to ensure that they converge to the same result at the expected Monte Carlo convergence rate ($N^{-0.5}$).
Figure~\ref{fig:moving_pellet} at bottom shows the norm converging at the expected rate.
This regressively verifies that delta tracking can be used in conjunction with continuously moving surfaces.
We also produced this same plot for both tracking methods using a collision estimator for both flux and fission reaction rates, all of which match the expected Monte Carlo convergence rate.

\section{Benchmark Problems}
\label{sec:benchmarks}

% vairiance reudciton and figure of merrit
%The performance of a given Monte Carlo algorithm for a specified problem is a function of solver variance ($\sigma^2$, from the Monte Carlo process itself) and the wall-clock runtime required to compute that solution.
%If a certain algorithm achieves a solution with fewer particles and reduced variance, as compared to another algorithm, it may still not increase solver efficiency if it takes too long to get that solution.
To measure performance, make comparisons, and draw conclusions of the various methods we use benchmark problems---accepted simulations in the radiation transport community.
A comparative measurement must be used to take into account both the variance and computation time of a given solution.
The figure of merit (FOM) is one such measure. In this work, we will use 
\begin{equation}
    \text{FOM} = \frac{1}{\hat{\sigma}^2 t_{wc}} \; ,
\end{equation}
where $\hat{\sigma}^2$ is the L$_{1}$ norm (over phase space) of the variance of the solution provided by the Monte Carlo solver and $t_{wc}$ is the measured wall-clock runtime the solver took to compute that solution.

% the problems we run themselves
We consider four time-dependent fixed-source benchmark problems, comparing the new voxelized tally scheme and hybrid methods proposed to surface tracking methods, on CPU and GPU machines.
Two problems are multi-group, and two are continuous energy.
Table~\ref{table:benchmark_problems} summarizes the size (in both mesh and particle count) of each benchmark problem.
Some problems may not be ideally suitable for delta tracking methods, but are used here to show both correctness under physical problem dynamics (e.g., moving surfaces) and demonstrate algorithmic performance under various physical parameters.
All problems are solved with at least four algorithms: surface tracking with a track length and collision estimator, and delta tracking with both a collision and track length estimator.
For the C5CE and modified Dragon burst problem, we also employ a hybrid delta-surface tracking algorithm using the collision and track length tally.

We explore these methods using time-dependent fixed-source problems.
``Fixed source'' and ``external source'' refer to a class of problems that are not eigenvalue problems.
In fixed source problems, the solution has a magnitude that depends on the prescribed source.
Neutron multiplication is allowed, provided the system is subcritical, and fixed source problems may or may not be time-dependent.
Eigenvalue ($k$ or $\alpha$) problems only involve sources that depend on the solution.  This work does not discuss how the implemented hybrid delta tracking methods impact $k$-eigenvalue or $\alpha$-eigenvalue calculation performance.

\begin{table}[htb]
  \centering
  \begin{tabular}{@{}l c c c @{}} \toprule
    Problem & $N_{\text{mesh}}$ & $N_{\text{particles}}$ & Energy physics \\ \midrule
    Kobayashi & \num{1.2e5} & \num{1e10} & MG (1 group) \\
    Modified Dragon Burst & \num{4.0e6} & \num{3e9} & CE (3 materials) \\
    Modified C5G7 & \num{3.9e6} & \num{1e7} & MG (7 groups, 8 delayed neutron precursor) \\
    C5CE & \num{5.4e5} & \num{1e5} & CE (7 materials) \\
    \bottomrule
  \end{tabular}
  \caption{Time-dependent benchmark problems, where ``MG'' indicates multi-group and ``CE'' continuous energy.} 
  \label{table:benchmark_problems} 
\end{table}

\subsection{Kobayashi problem}

% koby intro
We first consider a time-dependent version of the Kobayashi problem \cite{Kobayashi2001} introduced by Variansyah et al.~\cite{variansyah_mc23_mcdc} (a full description is provided by Variansyah~\cite{variansyah_2025_15069882}).
We use 10 batches with \num{e9} particles per batch (\num{e10} particles total) with surface tracking, standard delta tracking with a collision estimator, and a delta tracking using the voxelized tally method.
This problem contains two materials, a low mean-free-path region, characterizing a solid ($\Sigma=$\SI{0.1}{\per\centi\meter}), and a high mean-free-path region modeled by air, characterizing a void ($\Sigma=$\SI{1e-4}{\per\centi\meter}), in a single energy group.
There are \num{1.20e5} structured tally voxels in the $x$-$y$ plane and in time.
For this problem, we expect the voxelized tally method to perform better than standard delta tracking and surface tracking.

\begin{figure}[htbp]
    \centering
    \includegraphics[width=0.48\linewidth]{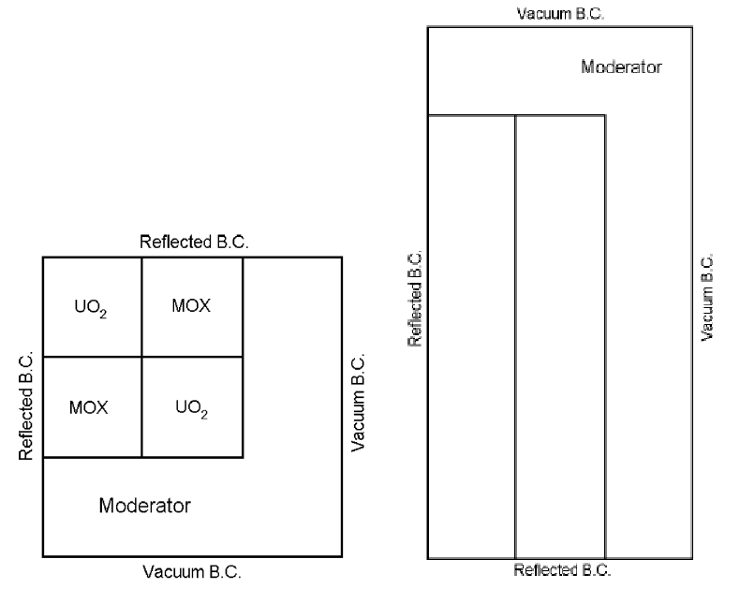}
    \includegraphics[width=0.48\linewidth]{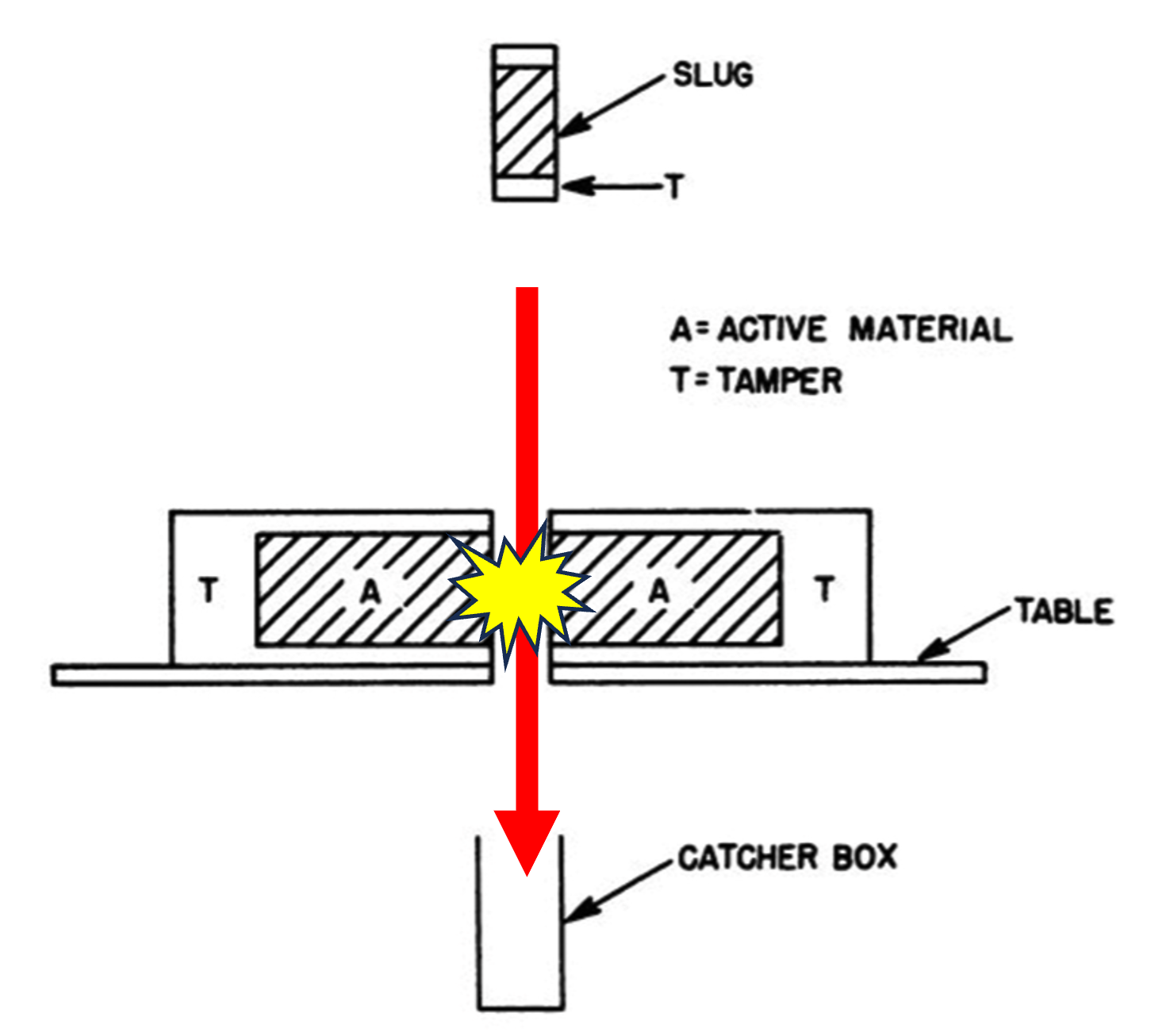}
    \caption{Schematics for benchmark problems: (left) C5G7 reactor quarter (via reflecting boundaries) \cite{hou2017oecd}, (right) Dragon burst problem \cite{kimpland_2021_dragon}.}
    \label{fig:schems}
\end{figure}

\subsection{Modified C5G7 problem}

\begin{figure}[htbp]
    \centering
    \includegraphics[width=.9\linewidth]{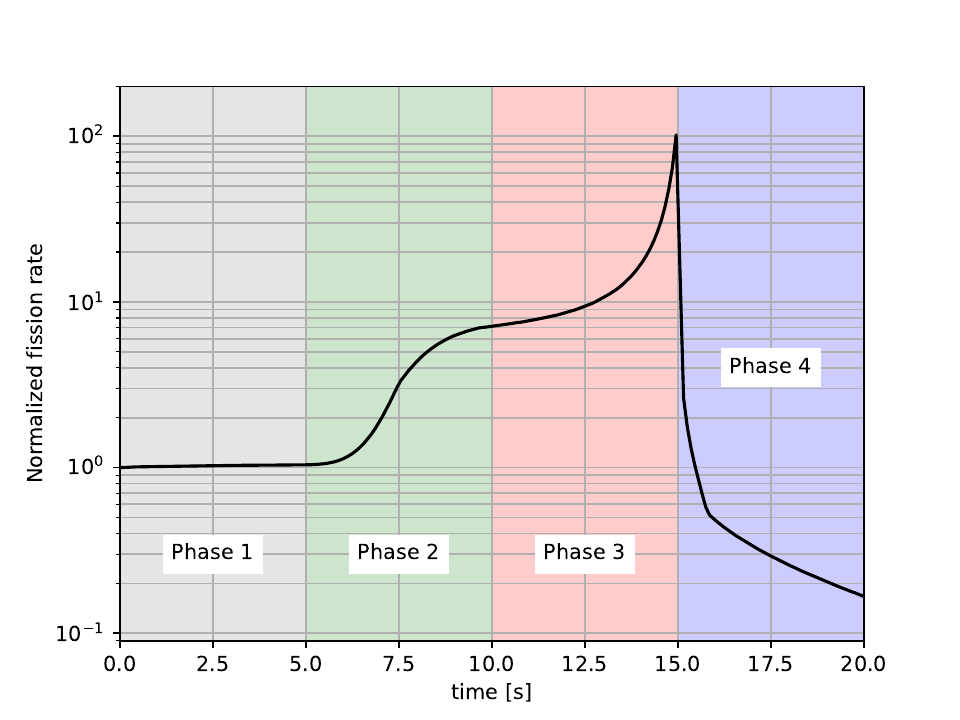}
    \includegraphics[width=.95\linewidth]{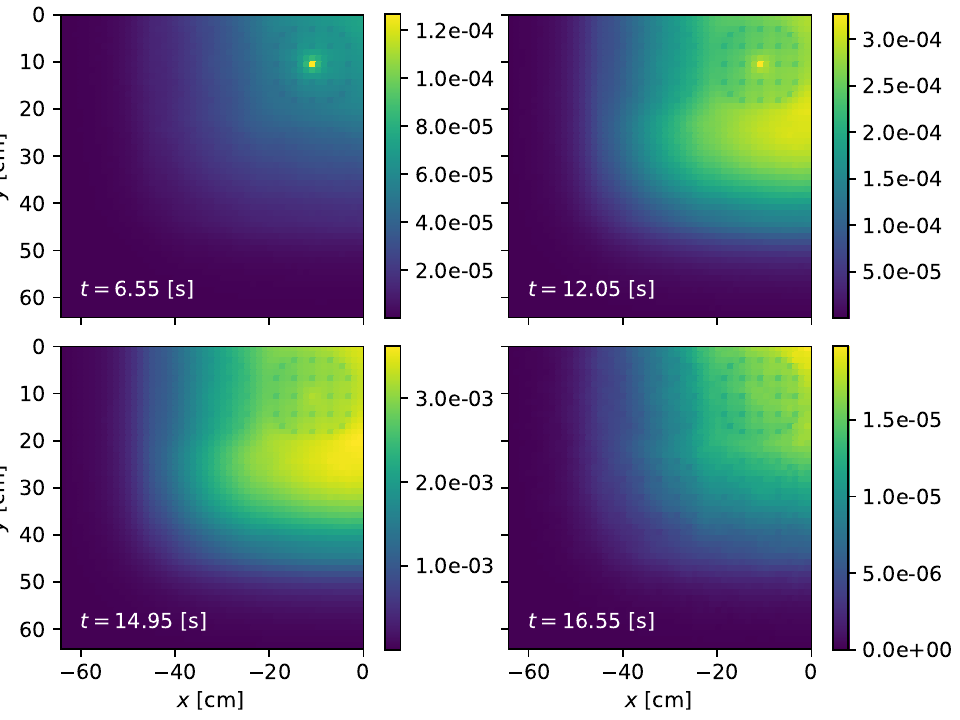}
    \caption{Modified C5G7 stuck rod accident simulation: (top) flux densities through time showing the four phases shaded as gray, green, red, and blue respectively; (bottom) scalar flux on $x$-$y$  plane (top view) at points in time.}
    \label{fig:c5g7}
\end{figure}

% acident intro
Figure~\ref{fig:schems} on the left shows the geometry for the next two simulations, which are based on the C5G7 benchmark problem \cite{hou2017oecd}.
In energy we use seven energy groups and eight delayed neutron precursor groups.
We model a four-phase accident in which a pressurized light water reactor increases in power via the removal of control rods
\cite{c5g7_zenodo}. The fixed-source source in the problem is located in throughout the active center pin of assembly four with a zero initial condition.
MC/DC does not include any treatments for cross section heating during a simulation so this kind of multi-physical calculation not considered in this work.
Figure~\ref{fig:c5g7} on the top shows the normalized flux density as a function of time (produced from point reactor kinetics) and the four phases shaded gray, green, red, and blue, respectively.
Phase one (shaded gray) starts with the reactor operating in steady state.
In phase two (shaded green), the control rods are removed from the reactor to cause the power to increase to a new steady state.
Phase three (shaded red) begins when a bank of control rods gets stuck in the fully withdrawn position.
Toward the end of phase three, the reactor sees a rapid increase in power that ends at \SI{15}{\s} when all control rods are forced into the reactor, ending the accident.
In the fourth and final phase, reactor power decays as the delayed neutron population dies out.
Figure \ref{fig:c5g7} on the bottom shows plots of average flux in the $x$-$y$ plane at various points in time, including at \SI{14.95}{\s}, the maximum power excursion.
These results are from a simulation with \num{e6} particles in \num{10} batches (a total of \num{e7} particles).

We first modeled this problem using the seven-group materials in the C5G7 benchmark description~\cite{hou2017oecd}, which we call C5G7 in the remainder of this work.
We use \num{3.9e6} mesh cells in a 3D, time- and energy-dependent tally mesh.
The movement of control rods into and out of the reactor is modeled with the continuous movement functionality of MC/DC \cite{variansyah_2023_highfidelity}.
To verify the delta tracking methods we explore convergence to the correct solutions using continuously moving surfaces, and we compare delta tracking solutions to solutions provided by surface tracking.
This further verifies that delta tracking can be used in conjunction with continuously moving surfaces.
Performance data is collected using \num{e6} particles in \num{10} batches (total of \num{e7} particles).

\subsection{C5CE problem}

Next, we define a continuous-energy version of the previous problem (named C5CE), in which the reactor undergoes the same four-phase accident.
We use the same C5G7 geometry \cite{c5g7_zenodo}, source distribution, and initial condition but establish each material using continuous energy cross sections.
Table~\ref{tab:c5ce} in Appendix~\ref{app:c5ce_mat} includes the material compositions for the C5CE problem.
Using this benchmark, we evaluate both the voxelized tally method and the hybrid-in-energy method described in Section~\ref{sec:cutoff}.
Figure~\ref{fig:majorant_c5ce} shows the macroscopic total and majorant cross sections as a function of energy for the materials in C5CE.
The neutron resolved-resonance region ends around \SI{10}{\kilo\electronvolt}, so we set the transition energy at \SI{50}{\kilo\electronvolt}.
Particles moving at energies above \SI{50}{\kilo\electronvolt} will use delta tracking, and those below will use surface tracking.

We expect the hybrid-in-energy approach to provide a significant speedup over the other methods explored in this problem, as it avoids the negative impacts of both tracking methods: surface tracking moving from surface to surface and the frequent rejection sampling associated with delta tracking. 
The C5CE problem includes the same continuously moving surfaces as C5G7, so it serves as an additional verification that delta tracking methods can be used in conjunction with continuously moving surfaces.
We model this problem with \num{5.44e5} mesh tally bins in a 2D $x$-$y$ geometry (integrated along $z$), time and energy-dependent tally mesh.
We use \num{1e5} particles in a single batch.

\subsection{Modified Dragon-burst problem}

% dragon intro

\begin{figure}[htbp]
    \centering
    \includegraphics[width=0.9\linewidth]{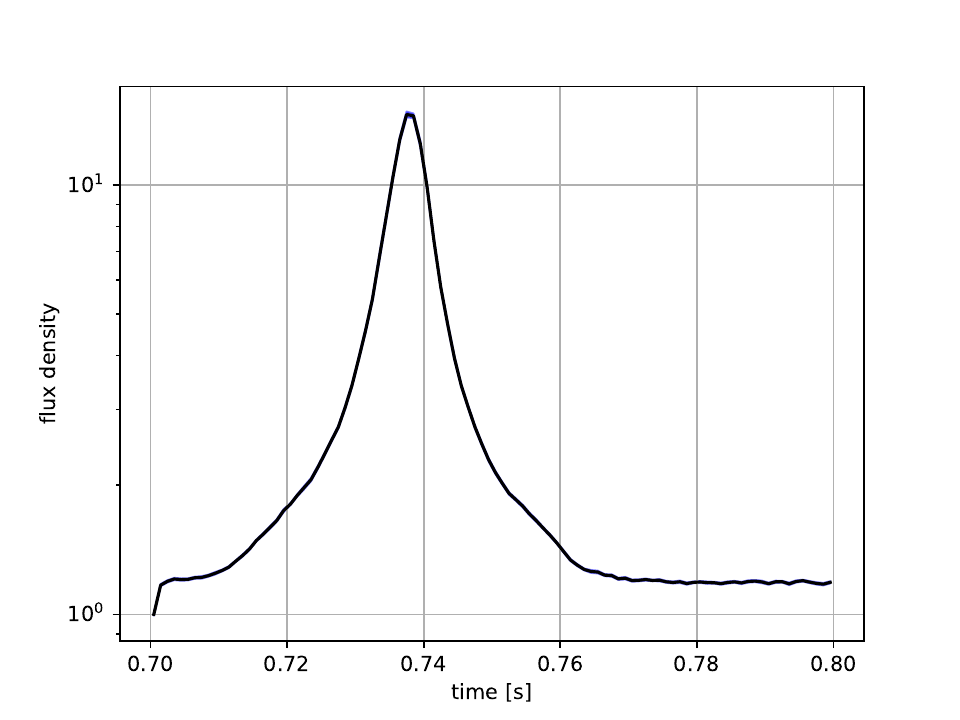}
    \includegraphics[width=0.95\linewidth]{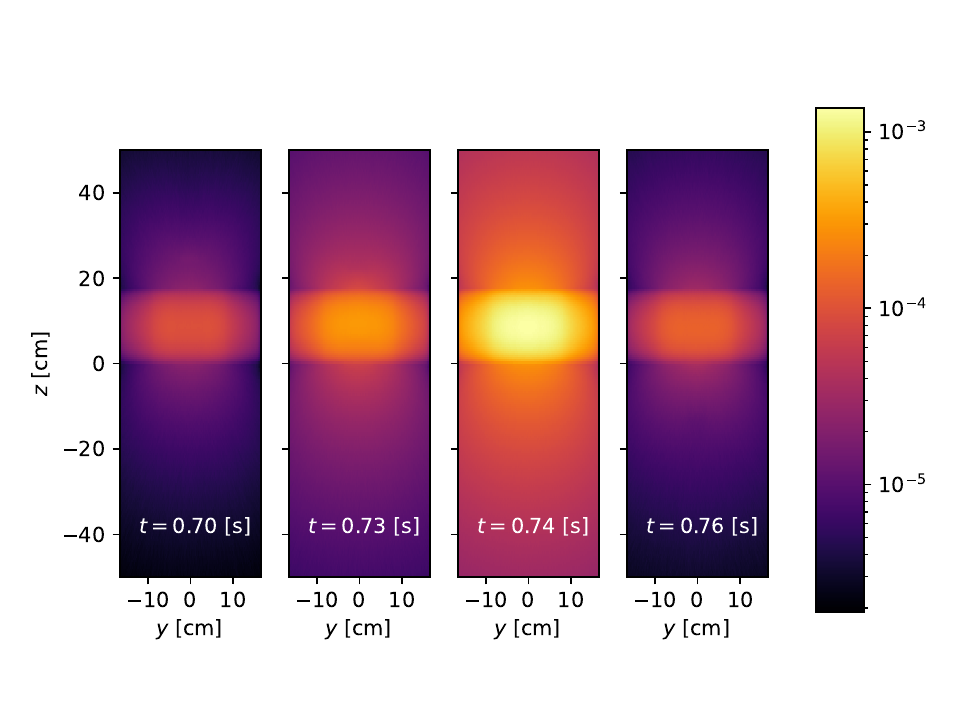}
    \caption{Modified Dragon burst simulation: (top) flux density as a function of time, (bottom) scalar flux on $y$-$z$ plane (side view) at points in time.}
    \label{fig:dragon_results}
\end{figure}

Conducted in 1944 during the Manhattan Project, the Dragon-burst experiments~\cite{kimpland_2021_dragon} proved that criticality and supercriticality could be achieved with prompt neutrons alone.
Previous experiments, such as the Chicago Pile One, depended on the presence of delayed neutrons to achieve criticality.
Figure~\ref{fig:schems} on the right shows a schematic of the experiment where a highly enriched (75\%) uranium hydride slug (UH$_{10}$) was dropped through a tamper with additional fuel.
The slug moves through the core, triggering a prompt supercritical reaction.
The reactivity excursion ends when gravity causes the slug to fall out of the core.
Kimpland et al.\ showed that this burst criticality experiment achieved an increase in the neutron population of more than nine orders of magnitude \cite{kimpland_2021_dragon}.
The total time of the reaction is fast enough such that cross section heating can be neglected \cite{kimpland_2021_dragon}.
Here we consider a less-reactive version of the Dragon burst problem (25\% enrichment) to test delta tracking with the hybrid-in-material method described in Section~\ref{sec:material_exc}, and to provide additional verification for delta tracking with continuously moving surfaces.
Figure~\ref{fig:dragon_results} shows the overall flux density as a function of time on the top and a $y$-$z$ plot of the scalar flux at various points in time on the bottom.

Figure \ref{fig:majorant_dragon} on the left shows the continuous energy total macroscopic cross sections in the model.
The majorant, tamper, and fuel cross sections are almost four orders of magnitude greater than the cross section of air.

We expect delta tracking algorithms to perform quite poorly in this model for a number of reasons.
First, when undergoing delta tracking, particles in the air region will experience a large number of rejected virtual collisions, since the majorant is orders of magnitude larger than the cross section of air.
Second, as much of the problem involves a near-void material, the collision estimator typically required by delta tracking will generate poor tallies in those regions.
Third, the problem is geometrically simple, consisting of a rectangular slug moving through a rectangular slab with a properly sized hole such that the slug can move through the slab.
The benefit of delta tracking over surface tracking is most pronounced in problems with many surface crossings \cite{woodcock_1965_deltatracking}.
Our objectives of comparing delta tracking methods for this specific problem are to confirm that delta tracking works with such a dynamic problem with continuously moving materials and to assess whether the hybrid-in-material method outperforms standard delta tracking.

\section{Results}

% TLEase add the following required packages to your document preamble:
% \usepackage{multirow}
\begin{table}
\centering
\begin{tabular}{@{}lccccc@{}} %lllllr
\toprule
Problem & Tracking Alg. & Estimator & Runtime [s] & $\left|\left|\hat{\sigma}^2\right|\right|_1$ & Figure of Merrit \\ \midrule
\multirow{4}{*}{Kobyashi}
 & surface  & TLE & \num{1574} & \num{4.514e-3} & 0.1407 \\
 & surface  & CE & \num{1014} & \num{9.552e-2} & 0.0103 \\
 & delta  & TLE & \num{1298} & \num{4.539e-3} & 0.1697 \\ 
 & delta  & CE & \num{817.2} & \num{2.062e-1}  & 0.0059 \\
 \midrule

% I just want to 
\multirow{4}{*}{C5G7}
 & surface  & TLE & \num{7926} & \num{1.295e-4} & \num{0.9744} \\
 & surface  & CE & \num{3173} & \num{1.295e-4} & \num{2.4336} \\
 & delta  & TLE & \num{2870.} & \num{1.400e-4} & \num{2.4889} \\
 & delta  & CE & \num{1820} & \num{2.152e-4} & \num{2.5515} \\
 \midrule

\multirow{6}{*}{C5CE} 
 & surface  & TLE & \num{734.6} & \num{1.812e-3} & \num{0.7484}\\
 & surface  & CE & \num{587.8} & \num{1.812e-3} & \num{0.9391}\\
 & delta  & TLE & \num{496.8} & \num{4.906e-3} & \num{0.4103} \\
 & delta  & CE & \num{555.0} & \num{1.173e-2} &  \num{0.1536} \\
 & hybrid-energy & TLE & \num{226.25} & \num{1.164e-3} & \num{3.798} \\ 
 & hybrid-energy & CE & \num{220.5} & \num{1.144e-3} & \num{3.965} \\ 
 \midrule

\multirow{4}{*}{Dragon} 
 & surface  & TLE & \num{3816} & \num{1.163e-6} & \num{221.4} \\
 & delta  & CE & DNF$^*$ & - & - \\
 & delta  & TLE & DNF$^*$ & - & - \\
 & hybrid-in-material & TLE & \num{15493} & \num{1.106e-6} & \num{58.40} \\
 \bottomrule
\end{tabular}
\caption{Results for benchmark problems on Dane (112$\times$ Intel Sapphire Rapids CPU cores). $^*$Did not finish in eight-hour time limit.}
\label{tab:dane_results}
\end{table}

%1.163327654958653e-06
%1.1056932706381267e-06

% \usepackage{multirow}
\begin{table}
\centering
\begin{tabular}{@{}lccccc@{}}
\toprule
Problem & Tracking Alg. & Estimator & Runtime [s] & $\left|\left|\hat{\sigma}^2\right|\right|_1$ & Figure of Merrit \\ \midrule
\multirow{4}{*}{Kobyashi} 
 & surface  & TLE & \num{973.9} & \num{4.514e-3} & 0.2275 \\
 & surface  & CE & \num{840.6} & \num{9.952e-2} & 0.0125 \\
 & delta  & TLE & \num{831.8} & \num{4.539e-3} & 0.2649 \\ 
 & delta  & CE & \num{620.7} & \num{2.062e-1}  & 0.0078 \\
 \midrule
 
\multirow{4}{*}{C5G7} 
 & surface  & TLE & \num{4598} & \num{1.305e-4} & \num{1.666} \\
 & surface  & CE & \num{2403} & \num{1.305e-4} & \num{3.1878} \\
 & delta  & TLE & \num{1615} & \num{1.400e-4} & \num{4.4205} \\ 
 & delta  & CE & \num{789.3} & \num{2.152e-4} & \num{5.8860} \\
 \midrule
 
\multirow{6}{*}{C5CE} 
 & surface  & TLE & \num{550.9} & \num{2.452e-3} & \num{0.7402}\\
 & surface  & CE & \num{465.9} & \num{2.452e-3} & \num{0.8753}\\
 & delta  & TLE & \num{500.7} & \num{1.690e-2} & \num{0.1182} \\
 & delta  & CE & \num{421.6} & \num{1.793e-2} &  \num{0.1323} \\
 & hybrid-energy & TLE & \num{291.7} & \num{1.557e-3} & \num{2.2023} \\ 
 & hybrid-energy & CE & \num{275.6} & \num{1.839e-3} & \num{1.9731} \\ 
 \midrule
 
\multirow{4}{*}{Dragon} 
 & surface & TLE & \num{3993} & \num{1.163e-06} & \num{2153} \\
 & delta & CE & DNF$^*$ & - & - \\
 & delta & TLE & DNF$^*$ & - & - \\
 & hybrid-material & TLE & DNF$^*$ & - & - \\ \bottomrule
\end{tabular}
\caption{Results for benchmark problems on Lassen (4$\times$ Nvidia Tesla V100). $^*$Did not finish in eight-hour time limit.}
\label{tab:lassen_results}
\end{table}

In this section, we discuss the performance results of the benchmarks modeled with surface, delta, and voxelized delta tracking, as well as the hybrid-in-energy and hybrid-in-material tracking algorithms.
We also verify that the various delta tracking methods converge to the same solution as surface tracking when transporting on a geometry with continuously moving surfaces.
All data, plots, the exact version of MC/DC, and input decks are available from Zenodo \cite{delta_restuls}.

\subsection{Performance results}

% testing systems
All results presented below are executed on the Dane and Lassen high-performance computing systems at Lawrence Livermore National Laboratory (LLNL).
Dane is a CPU-only system with dual-socket Intel Xeon Sapphire Rapids CPUs, each with 56 cores for a total of 112 CPU cores per node. 
Lassen is the open collaboration sibling to the Sierra machine with four Nvidia Tesla V100 GPUs and two IBM Power 9 CPUs per node.
We make all performance statements with respect to a whole node of Dane (112 MPI threads, CPU) and a whole node of Lassen (4 MPI threads, GPU).
We compile to CPUs with Numba v0.60.0, and compile to Nvidia GPUs with CUDA v11.8 and Nvidia-PTX with Numba v0.59.0\footnote{Numba v0.59.1 is the most recent version to support Power9 CPUs.}.
The delta tracking methods are built using MC/DC v0.12.0 \cite{transport_cement_mcdc_2024}, compiling on GPUs with Harmonize v0.0.2 \cite{harmonize}.
All floating-point operations use double precision.

% Introduce the table
Tables~\ref{tab:dane_results} and \ref{tab:lassen_results} show the wall clock runtimes, L$_1$ norm of the Monte Carlo variance ($\left|\left|\hat{\sigma}^2\right|\right|_1$), and figures of merit (FoM) for benchmark problems using various transport algorithms and flux estimators on the Dane and Lassen machines, respectively.

\begin{figure}[htbp]
    \centering
    \includegraphics[width=\textwidth]{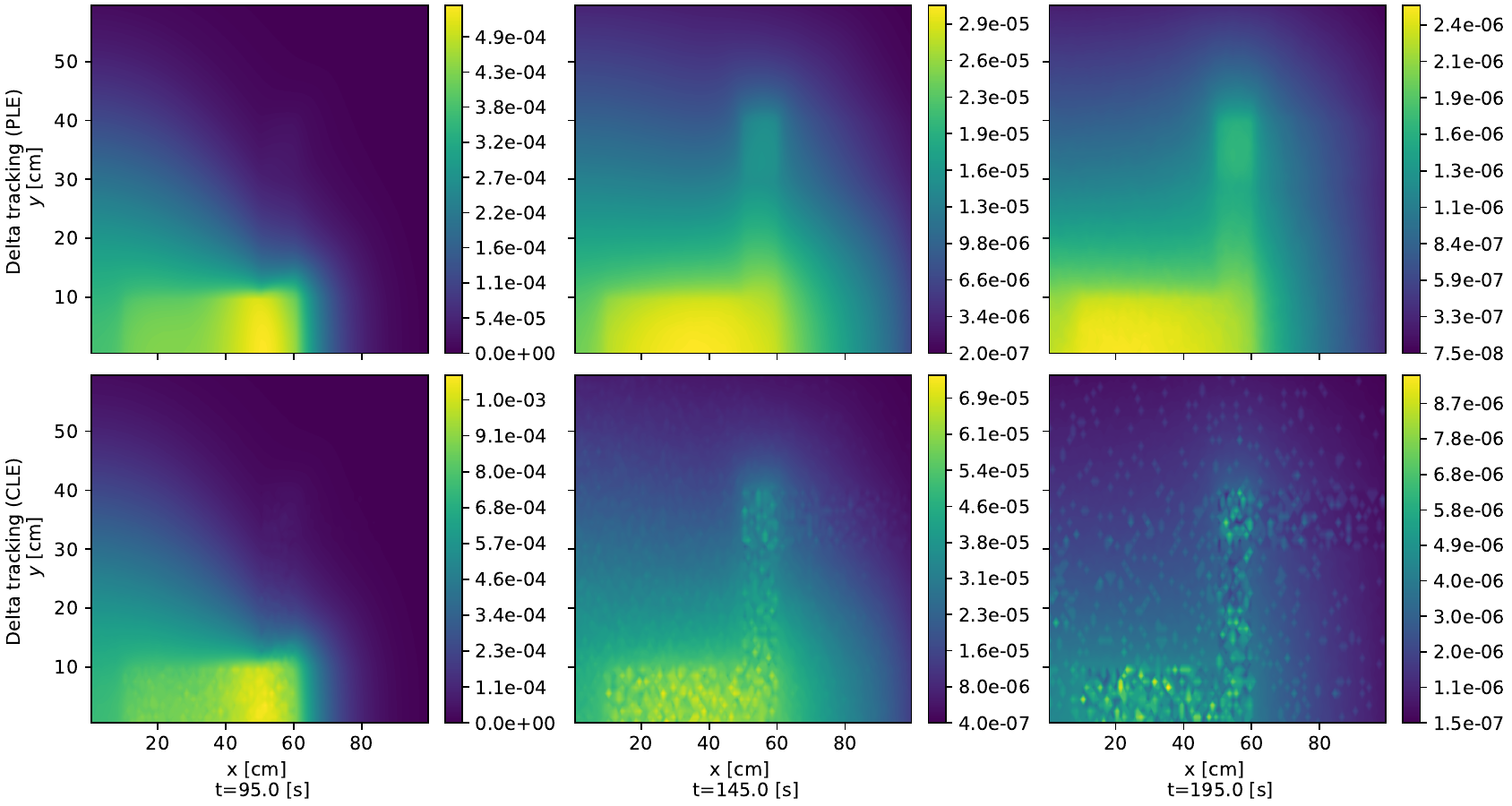}
    \caption{Comparison of delta tracking using the track length estimator (top row) and collision estimator (bottom row) at three points in time. The solution produced with the collision estimator has much higher variance.}
    \label{fig:koby}
\end{figure}

% Results Koby
On Dane, delta tracking and a collision estimator dramatically improve runtime for the Kobayashi problem compared with surface tracking, with a collision or track length estimator.
However, this speedup does not compensate for the two orders of magnitude additional variance on the tallies of interest incurred by the collision estimator.
This leads to a significantly smaller figure of merit for standard delta tracking and surface tracking (using a collision estimator) compared with results from the track length estimator.
Figure~\ref{fig:koby} shows the simulated flux at various points in time for the Kobayashi problem computed using standard delta tracking with a collision estimator (bottom) and voxelized delta tracking (top).
The collision estimator yields a solution with a higher variance (appearing as static) result compared to that of the track length estimator.
The voxelized tally method achieves a 21\% decrease in wall clock time for the same amount of normed variance, leading to a moderately improved figure of merit (\num{0.1697} and \num{0.1407} for the voxelized method and surface tracking, respectively).
On GPUs, the pattern is the same, with the voxelized method performing slightly better than surface tracking with a track length estimator.
All algorithms run between 1.2$\times$ and 1.7$\times$ faster on Lassen than Dane. 

%This pattern of results, where traditional delta tracking is the fastest wall-clock runtime with a normalized variance orders of magnitude above methods using a track length estimator, will persist through the analysis of the other problems we consider.
%As will the behavior of the voxelized delta method's runtime sitting between surface tracking and traditional delta tracking with similar variance to surface tracking results.

% Restuls C5G7
For modified C5G7 on Dane, standard delta tracking performs the best with similar L1 norm of variance (\SI{1820}{\s} and \num{2.152e-4}, respectively).
Surface tracking sees the longest runtime and voxelized delta tracking sits between the two (\SI{7926}{\s} and \SI{2870}{\s}, respectively) with roughly the same error (~\num{1.3e-4}).
For C5G7, there does not appear to be a benefit in using the track length estimator.
This pattern is also observed in the Lassen results, with between \num{1.7}$\times$ and \num{2}$\times$ wall clock runtime speedup for all tracking methods when moving from Dane to Lassen, while the variance remains about the same.
Voxelized delta tracking shows a \num{2.6}$\times$ and \num{2.7}$\times$ higher figure of merit over surface tracking on Dane and Lassen, respectively, and standard delta tracking performs slightly better than the voxelized tally method.

% Restuls C5CE
In solving the C5CE problem, surface tracking methods yielded the highest figures of merit, with the collision and track length estimator.
The speedup of Lassen over Dane is now lower (between \num{0.9}$\times$ and \num{1.3}$\times$), indicating improvements are needed for our continuous energy physics when implemented on GPUs.
The hybrid-in-energy method provides significant improvement in FoM with an order of magnitude increase: \num{11}$\times$ on Dane, \num{7}$\times$ on Lassen. 
The use of the collision estimator in the hybrid method slightly outperforms the use of the track length estimator.

% Restuls Dragon
The performance of the methods on the modified Dragon problem is at odds with observations of the algorithms on the other problems.
Delta tracking simulations (both standard and voxelized) do not reach the end of the specified end time on Dane or Lassen in the eight hours allocated, on either machine.
On Dane, as predicted, the hybrid-in-material method improves performance over the other two methods, as it completes the simulation in about four hours. 
However, standard surface tracking requires only one hour for this simulation.
The hybrid-in-material method did not finish on Lassen.
These observations are consistent with the understanding that delta tracking is typically less effective for problems with fewer surfaces and significant material heterogeneities.

\section{Discussion, Conclusions, and Future Work}
\label{disucssions}

We have implemented delta tracking in MC/DC on CPUs and GPUs, including using a voxelized track length tally to efficiently score scalar flux.
We verify the solution produced by this method against various steady-state and time-dependent analytic benchmark problems available in the MC/DC verification suite.
We also verify that delta tracking functions properly with continuously moving surfaces in MC/DC using the fuel pellet problem.
Figures of merit improve modestly in large-scale multi-group and continuous energy time-dependent benchmark problems when using this tracking and tallying technique on CPUs and GPUs.

We have also demonstrated a novel hybrid-in-energy surface-delta tracking scheme, where surface and delta tracking are used at low and high energies, respectively.
On the continuous energy version of the C5G7 benchmark geometry undergoing a four-phase transient, we observe an order of magnitude increase in FOM with the hybrid-in-energy scheme, on both CPU and GPU nodes.
We have also confirmed that hybrid surface-delta tracking methods can yield improved results in problems with significant void regions \cite{leppanen_2010_burnup}.

An efficient Monte Carlo algorithm is a combination of a numerical method that obtains a lower variance with fewer particles and an efficient implementation of that numerical method.
What makes a given delta or surface tracking algorithm more or less ``efficient'' can vary code-to-code depending on the optimizations software developers have chosen.
Not every code implements tallying the same way as MC/DC, meaning the added efficiency of using the voxelized tally may not be a viable option in other Monte Carlo neutron transport codes.

The main takeaway of this work is that delta and surface tracking do not have to be treated as discrete choices in the numerical method, as corroborated by papers from the developers of Serpent2 and MONK/MCBEND~ \cite{leppanen_2017_collision, richards_monk_2015}.
Greater performance can potentially be achieved by mixing and matching the underlying transport methods, given the physical parameters of a simulation and the relative strengths and weaknesses of a given transport application.
As discussed in Section \ref{sec:implementation}, implementing delta tracking in a surface tracking code is relatively simple, given a method to calculate a macroscopic majorant cross section.
%This process has been completed at least twice now in MCATK \cite{morgan_2021_mcatk} and MC/DC with work ongoing in OpenMC and MCNP. % cite their branch

The use of the voxelized tally method in void regions is promising, but the lack of reaction rate tallies limits its applicability.
Work is ongoing to produce efficient methods of computing relevant macroscopic cross sections defined on a structured mesh while a particle is undergoing transport.
This process is straightforward for multi-group cross sections where reaction rates can be determined entirely via post-processing, but we are working to identify the most efficient method for continuous energy transport.

MC/DC does not currently treat the unresolved resonance region.
Work is ongoing to support unresolved resonances; when implemented the energy cutoff for the hybrid-in-energy delta tracking method could use the average value of the cross section to determine a cutoff value.
For the majorant determination in the unresolved resonance region, the maximum cross section value in the unresolved resonance table will be used.
This work only includes analysis for fixed source time-dependent problems.
We leave the analysis of $k$ or $\alpha$ eigenvalue calculations for future work.

Methods for reducing the dimensionality of the majorant are also being explored to achieve a more-efficient majorant cross section lookup in distance to collision operations.
Experiments with the collision estimator for scalar flux and methods of tallying into vacuum and low interaction rate regions are being considered, including implementing a cutoff method \cite{leppanen_2010_burnup} or weighted delta tracking \cite{morgan_weighted-delta-tracking_2015}.
%Research is also ongoing into methods to tally with multiple estimators .
%If only one estimator is used at a time then we avoid the need for complex covariance computations (e.g., those implemented in MCNP \cite{urbatsch_estimation_1995, MCNP_RisingArmstrongEtAl}).

%Combinations of the delta and surface tracking algorithms proves to be a compiling field of research in Monte Carlo neutron transport.
%The combination of these two tracking algorithms allows for greater performance adaptability on a broader set of problem physics where one scheme may out perform another.
%Either method is a valid sampling of the cumulative probability distribution function at any point in transport.
%Modest to significant improvements may occur when taking advantage of that fact.

\section*{Acknowledgments}
We thank Patrick Shriwise and Paul Romano of Argonne National Laboratory, Mike Rising of Los Alamos National Laboratory, Jaakko Leppänen of VTT Technical Research Center of Finland, and Simon Richards of ANSWERS software for productive conversations.
We also thank the high performance computing staff at Lawrence Livermore National Laboratory for continued support using the Dane machine. 

This work was supported by the Center for Exascale Monte-Carlo Neutron Transport (CEMeNT) a PSAAP-III project funded by the Department of Energy, grant number: DE-NA003967.

\bibliographystyle{IEEEtranDOIandURLwithDate}
\bibliography{main}
\newpage

\appendix
\section{Continuous Energy Macroscopic Majorant}
\label{app:majorant}

To compute a unified energy grid per material, we combine the energy grids from all nuclides in a given material. 
Then we call \texttt{numpy.unique()}, which returns a sorted array (from smallest to largest) with no repeating elements \cite{van_der_walt_numpy_2011}.
To compute a unified energy grid for the whole problem, we do the same but with all the nuclides in the entire problem.

Computing a macroscopic majorant for nuclides that are not on a unified energy grid requires two levels of interpolation to put a given macroscopic total cross section on a unified energy grid.
First, interpolae from each nuclide's microscopic cross section onto a material's unified energy grid to compute a macroscopic total cross section.
Then, perform a second interpolation from the material to the majorant's unified energy grid.
We use \texttt{scipy.interpolate.interp1d()} to interpolate from one energy grid to the next \cite{2020SciPy-NMeth}. 

The following code shows how this is done in MC/DC:

\begin{lstlisting}[language=Python]
import scipy
import numpy

# unify the energy grids from all nuclides
majorant_energy_grid = np.array([])
for n in range(N_nuclide):
    nuclide = mcdc['nuclides'][n]
    majorant_energy_grid = np.append(majorant_energy_grid, nuclide['E_xs'])

# sort energy grid and eliminate duplicate points
majorant_energy_grid = np.unique(majorant_energy_grid)
majorant_xsec = np.zeros_like(majorant_energy_grid)

for m in range(N_material):

    material = mcdc['materials'][m]

    material_energy_grid = np.array([])

    # copmute a unified energy grid across all nuclides of a given material
    for n in range(material['N_nuclide']):
        nuclide = mcdc['nuclides'][n]
        material_energy_grid = np.append(
            material_energy_grid, nuclide['E_xs']
        )
    material_energy_grid = np.unique(material_energy_grid)
    MacroXS = np.zeros_like(material_energy_grid)

    # compute the macroscopic total cross section of a material on its unified
    # energy grid
    for n in range(material['N_nuclide']):
        ID_nuclide = material['nuclide_IDs'][n]
        nuclide = mcdc['nuclides'][ID_nuclide]

        # Get nuclide density
        N = material['nuclide_densities'][n]

        # putting the microscopic cross-sections on the unifed
        # material energy grid
        total_micro_xsec_unified = scipy.interpolate.interp1d(
            nuclide['E_xs'], nuclide['ce_total'], bounds_error=False
        )
        total_micro_xsec_unified = total_micro_xsec_unified(
            material_energy_grid
        )

        # Accumulate
        MacroXS += N * total_micro_xsec_unified

    # puting the total macroscopic cross sections on on the majorant energy grid
    total_xsec_unified = scipy.interpolate.interp1d(
        material_energy_grid, MacroXS, bounds_error=False
    )
    total_xsec_unified = total_xsec_unified(majorant_energy_grid)

    # compares old majorant xsec and the currently evaluated unified xsec 
    # and picks the larger xsecs
    majorant_xsec = np.max((majorant_xsec, total_xsec_unified), axis=0)

\end{lstlisting}

This process results in a large majorant cross-section that is very large.
Other, more efficient algorithms exist to produce a similarly accurate majorant with fewer points.
Delta tracking codes like Serpent2, GUARDYAN, and IMPC avoid the need for energy grid unification by having all nuclides on a unified energy grid in their data library \cite{leppanen_2015_serpent, molnar_guardyan_2019, fang_development_2022}.

\newpage

\section{C5CE Material Definition}
\label{app:c5ce_mat}
%\vspace{-1em}
\begin{longtable}{@{}l l l@{}}
\toprule
Material                         & Nuclide & Atom fraction          \\ \midrule
\endfirsthead
\endhead
\multirow{5}{*}{UO$_2$ Fuel}        & O-16     & \num{0.04585265389377734}    \\ 
                                 & O-17     & \num{1.7419604031574338e-05} \\ %\cline{2-3} 
                                 & O-18     & \num{9.19424166352541e-05}   \\ %\cline{2-3} 
                                 & U-235    & \num{0.0007217486041189947}  \\ %\cline{2-3} 
                                 & U-238    & \num{0.02224950230720295}    \\ \midrule
                                 & O-17     & \num{1.743649552488715e-05}  \\ %\cline{2-3} 
\multirow{5}{*}{MOX-43 Fuel}     & O-16     & \num{0.04589711643122753}    \\ %\cline{2-3} 
                                 & O-17     & \num{1.743649552488715e-05}  \\ %\cline{2-3} 
                                 & O-18     & \num{9.203157163056531e-05}  \\ %\cline{2-3} 
                                 & U-235    & \num{0.0003750264168772414}  \\ %\cline{2-3} 
                                 & U-238    & \num{0.02262319599228636}    \\ \midrule
\multirow{5}{*}{MOX-7 Fuel}      & O-16     & \num{0.04583036614158277}    \\ %\cline{2-3} 
                                 & O-17     & \num{1.741113682662514e-05}  \\ %\cline{2-3} 
                                 & O-18     & \num{9.189772587857765e-05}  \\ %\cline{2-3} 
                                 & U-235    & \num{0.0005581382302893396}  \\ %\cline{2-3} 
                                 & U-238    & \num{0.022404154012604437}   \\ \midrule
\multirow{5}{*}{MOX-87 Fuel}     & O-16     & \num{0.04585265389377734}    \\ %\cline{2-3} 
                                 & O-17     & \num{1.7419604031574338e-05} \\ %\cline{2-3} 
                                 & O-18     & \num{9.19424166352541e-05}   \\ %\cline{2-3} 
                                 & U-235    & \num{0.0007217486041189947}  \\ %\cline{2-3} 
                                 & U-238    & \num{0.02224950230720295}    \\ \midrule
\multirow{5}{*}{Guide Tube}      & H-1      & \num{0.050347844752850625}   \\ %\cline{2-3} 
                                 & H-2      & \num{7.842394716362082e-06}  \\ %\cline{2-3} 
                                 & O-16     & \num{0.025117935412784034}   \\ %\cline{2-3} 
                                 & O-17     & \num{9.542402714463945e-06}  \\ %\cline{2-3} 
                                 & O-18     & \num{5.03657582849965e-05}   \\ \midrule
\multirow{5}{*}{Fission Chamber} & H-1      & \num{0.050347844752850625}   \\ %\cline{2-3} 
                                 & H-2      & \num{7.842394716362082e-06}  \\ %\cline{2-3} 
                                 & O-16     & \num{0.025117935412784034}   \\ %\cline{2-3} 
                                 & O-17     & \num{9.542402714463945e-06}  \\ %\cline{2-3} 
                                 & O-18     & \num{5.03657582849965e-05}   \\ \midrule
\multirow{12}{*}{Control Rod}    & Ag-107   & \num{0.023523285675833942}   \\ %\cline{2-3} 
                                 & Ag-109   & \num{0.02185429814297804}    \\ %\cline{2-3} 
                                 & In-113   & \num{0.0003421922042655644}  \\ %\cline{2-3} 
                                 & In-115   & \num{0.007651085167039375}   \\ %\cline{2-3} 
                                 & Cd-106   & \num{3.38816276451386e-05}   \\ %\cline{2-3} 
                                 & Cd-108   & \num{2.4166172970990425e-05} \\ %\cline{2-3} 
                                 & Cd-110   & \num{0.0003393605596264083}  \\ %\cline{2-3} 
                                 & Cd-111   & \num{0.0003482051612205208}  \\ %\cline{2-3} 
                                 & Cd-112   & \num{0.0006561061533306398}  \\ %\cline{2-3} 
                                 & Cd-113   & \num{0.00033274751904988726} \\ %\cline{2-3} 
                                 & Cd-114   & \num{0.0007825159207295705}  \\ %\cline{2-3} 
                                 & Cd-116   & \num{0.00020443276053837845} \\ \midrule
\multirow{5}{*}{Moderator}       & H-1      & \num{0.050347844752850625}   \\ %\cline{2-3} 
                                 & H-2      & \num{7.842394716362082e-06}  \\ %\cline{2-3} 
                                 & O-16     & \num{0.025117935412784034}   \\ %\cline{2-3} 
                                 & O-17     & \num{9.542402714463945e-06}  \\ %\cline{2-3} 
                                 & O-18     & \num{5.03657582849965e-05}   \\ \bottomrule
\caption{Materials used in C5CE problem}
\label{tab:c5ce}\\
\end{longtable}

\end{document}